\documentclass[%
 reprint,
 superscriptaddress,
 nofootinbib,
 amsmath,amssymb,
 aps,
]{revtex4-2}

\usepackage{graphicx}% Include figure files
\usepackage{dcolumn}% Align table columns on decimal point
\usepackage{bm}% bold math
\usepackage{hyperref}% add hypertext capabilities
\usepackage{multirow}
\usepackage[table]{xcolor}

\newcommand{\mnras}{MNRAS}
\newcommand{\apjl}{ApJL}
\newcommand{\aap}{A\&A}
\newcommand{\jcap}{JCAP}

\newcommand\pN{\mathcal{N}}

\newcommand\pH{\mathcal{H}}

\newcommand{\Nbar}{\ensuremath{\bar{N}}}
\newcommand{\hubble}{\ensuremath{H_0}}
\newcommand{\tradH}{\ensuremath{\hat{H}_0^{\rm trad}}}
\newcommand{\LFIH}{\ensuremath{\hat{H}_0^{\rm LFI}}}
\newcommand{\hunits}{\ensuremath{\mathrm{km\, s^{-1}\,Mpc^{-1}}}}
\newcommand{\dec}{\ensuremath{q_0}}
\newcommand{\tradq}{\ensuremath{\hat{q}_0^{\rm trad}}}
\newcommand{\LFIq}{\ensuremath{\hat{q}_0^{\rm LFI}}}
\newcommand{\cosmo}{\ensuremath{\boldsymbol{\Omega}}}
\newcommand{\biasH}{\ensuremath{b_{H_0}}}
\newcommand{\sbiasH}{\ensuremath{\sigma_{b_{H_0}}}}

\newcommand{\biasq}{\ensuremath{b_{q_0}}}

\newcommand{\fsH}{\ensuremath{f_{\sigma}^{H_0}}}

\newcommand{\fsq}{\ensuremath{f_{\sigma}^{q_0}}}

\newcommand{\incr}{\ensuremath{\%\hat{\sigma}_{incr}^{\hubble}}}

\newcommand{\lo}{\ensuremath{\lambda_1}}
\newcommand{\lt}{\ensuremath{\lambda_2}}
\newcommand{\nt}{\ensuremath{n_{\rm train}}}
\newcommand{\nv}{\ensuremath{n_{\rm val}}}
\newcommand{\nb}{\ensuremath{n_{\rm batch}}}

\newcommand{\zmax}{\ensuremath{z_{\rm{max}}}}
\newcommand{\zi}{\ensuremath{z_i}}
\newcommand{\hz}{\ensuremath{\hat{z}}}
\newcommand{\obszi}{\ensuremath{\hat{z}_i}}
\newcommand{\obsz}{\ensuremath{\hat{\boldsymbol{z}}}}
\newcommand{\di}{\ensuremath{d_i}}
\newcommand{\hd}{\ensuremath{\hat{d}}}
\newcommand{\obsdi}{\ensuremath{\hat{d}_i}}
\newcommand{\obsd}{\ensuremath{\hat{\boldsymbol{d}}}}
\newcommand{\vi}{\ensuremath{v_i}}
\newcommand{\vunits}{\ensuremath{\mathrm{km\; s^{-1}}}}
\newcommand{\hv}{\ensuremath{\hat{v}}}
\newcommand{\obsvi}{\ensuremath{\hat{v}_i}}
\newcommand{\obsv}{\ensuremath{\hat{\boldsymbol{v}}}}
\newcommand{\obsx}{\ensuremath{\hat{\boldsymbol{x}}}}
\newcommand{\obst}{\ensuremath{\hat{\boldsymbol{t}}}}
\newcommand{\thr}{\ensuremath{\rho_{*}}}

\usepackage{physics}

\definecolor{Gray}{gray}{0.9}

\begin{document}

\preprint{APS/123-QED}

\title{Unbiased likelihood-free inference of the Hubble constant from light standard sirens}
%\title{Likelihood-free inference of the Hubble constant from standard sirens in the presence of selection effects}
%\title{Likelihood-free inference of the Hubble constant from selected light standard sirens}

\author{Francesca Gerardi}
 \email{francesca.gerardi.19@ucl.ac.uk} %Lines break automatically or can be forced with \\
\author{Stephen M. Feeney}%
\affiliation{%
 Department of Physics \& Astronomy, University College London, Gower Street, London WC1E 6BT, UK\\
}%

\author{Justin Alsing}
\affiliation{
 Oskar Klein Centre for Cosmoparticle Physics, Department of Physics, Stockholm University, Stockholm SE-106 91, Sweden\\
}%

\date{\today}% It is always \today, today,

\begin{abstract}

Multi-messenger observations of binary neutron star mergers offer a promising path towards resolution of the Hubble constant ($\hubble$) tension, provided their constraints are shown to be free from systematics such as the Malmquist bias. In the traditional Bayesian framework, accounting for selection effects in the likelihood requires calculation of the expected number (or fraction) of detections as a function of the parameters describing the population and cosmology; a potentially costly and/or inaccurate process. This calculation can, however, be bypassed completely by performing the inference in a framework in which the likelihood is never explicitly calculated, but instead fit using forward simulations of the data, which naturally include the selection. This is Likelihood-Free Inference (LFI). Here, we use density-estimation LFI, coupled to neural-network-based data compression, to infer $\hubble$ from mock catalogues of binary neutron star mergers, given noisy redshift, distance and peculiar velocity estimates for each object. We demonstrate that LFI yields statistically unbiased estimates of $\hubble$ in the presence of selection effects, with precision matching that of sampling the full Bayesian hierarchical model. Marginalizing over the bias increases the $\hubble$ uncertainty by only $6\%$ for training sets consisting of $O(10^4)$ populations. The resulting LFI framework is applicable to population-level inference problems with selection effects across astrophysics.

%\begin{description}
%\end{description}
\end{abstract}

\maketitle

%\tableofcontents

\section{INTRODUCTION}

In recent years, late-time measurements \cite{Riess:2020, Birrer:2018vtm, Wong:2019kwg} of the Hubble Constant, $H_{0}$, have diverged from estimates provided by early-time probes~\cite{Planck2018, Addison_etal:2018, DES_SPT:2018, Philcox_etal:2020} (see Refs.~\cite{Bernal:2016gxb,Verde:2019ivm,Bernal:2021yli} for a summary). At the heart of the discrepancy is a $4.2\sigma$ tension between the latest direct measurement of $\hubble = (73.2 \pm 1.3)\;\hunits$ by the SH0ES Team's Cepheid-supernova distance ladder~\cite{Riess:2020} and the model-dependent value of $\hubble = (67.4 \pm 0.5)\;\hunits$ inferred from observations of the cosmic microwave background (CMB) anisotropies by the Planck satellite \cite{Planck2018}. While unforeseen systematic effects \cite{Rigault_2015,2015ApJ...812...31J,Rigault:2018ffm,2018ApJ...867..108J,Freedman:2020dne,Brout:2020msh} might be the cause of this disagreement, it is possible that this is a hint for new physics beyond the standard $\Lambda$CDM model (see Ref.~\cite{DiValentino:2021izs} for a comprehensive summary of potential theoretical solutions). Despite considerable effort, however, no consensus on an explanation has been reached. This strongly motivates the need for a new, independent, direct probe of $\hubble$. Gravitational waves (GWs) emitted by compact-object mergers -- so-called \textit{standard sirens} -- are very promising in this regard \cite{Schutz,HolzandHughes,Dalal:2006qt,Nissanke:2009kt,PhysRevD.85.023535,PhysRevLett.108.091101,Nissanke:2013fka,PhysRevD.93.083511,PhysRevD.95.043502,PhysRevLett.121.021303,10.1093/mnras/sty090,Feeney_BNSpredictions,Vitale:2018wlg,Gray:2019ksv,Feeney:2020kxk,Vitale:2020aaz}, since their amplitude provides a self-calibrated estimate of the luminosity distance, $d$, depending only on General Relativity. 

There are three types of compact-object systems typically considered for $\hubble$ studies \cite{prospects}: binary black holes (BBH), binary neutron stars (BNS) and neutron star - black hole (NSBH) systems. The potential for BNS and NSBH systems to have electromagnetic (EM) counterparts makes them particularly promising, as if an EM counterpart can be detected, the merger's host galaxy can be identified and its redshift measured, yielding $\hubble$ when combined with $d$ \cite{Dalal:2006qt,Nissanke:2009kt,Nissanke:2013fka,Vitale:2018wlg,Chen_BNSpredictions,Feeney_BNSpredictions,2018MNRAS.475.4133S,Abbott_H0infer}.
The first BNS system detected by the LIGO-Virgo Consortium, GW170817~\cite{TheLIGOScientific:2017qsa}, also produced an EM counterpart~\cite{GBM:2017lvd}, constraining $\hubble$ to $70.0^{+12.0}_{-8.0}\;\hunits$ \cite{Abbott_H0infer}. The 10\% constraints produced by this single event are expected to shrink to $\sim 1\%$ in the next 5-10 years once $O(100)$ events have been observed~\cite{Chen_BNSpredictions,Feeney_BNSpredictions,2018MNRAS.475.4133S}.

For standard siren estimates of $\hubble$ to resolve the current tension, they must be shown to be free from systematic errors. Standard siren datasets suffer from Malmquist bias~\cite{1922MeLuF.100....1M,1925MeLuF.106....1M} which, left untreated, results in $\hubble$ being overestimated. Traditional Bayesian methods must therefore take this effect into account by including in the likelihood terms involving the number (or, equivalently, fraction) of mergers that are expected to be detected given a set of population and cosmological parameters, $\Nbar(\cosmo)$~\cite{Loredo:2004nn,Abbott_H0infer,Mandel:2018mve,Mortlock_etal:2019,Vitale:2020aaz}. The simplest method for calculating the expected number of detections is through Monte Carlo integration, i.e., repeated simulations of the dataset. Implementing this directly within a posterior sampling algorithm is, however, completely unfeasible, given the sheer number of simulations that would be needed. Instead, a single large catalogue of detected mergers can be generated using a fiducial set of population parameters and then reweighted to approximate $\Nbar$ for any value of population parameters sampled~\cite{Tiwari:2017ndi}. If the distribution of object parameters changes rapidly as a function of population parameters, however, a large (potentially computationally unfeasible) number of fiducial-population simulations are required to guarantee there are enough non-zero weights for the estimate of $\Nbar$ to be reliable (the \textit{effective} number of detected mergers must be at least four times the measured number~\cite{2019RNAAS...3...66F}). Alternatively, $\Nbar$ can be evaluated on a grid of $\cosmo$ and interpolated to generic population parameters~\cite{Mortlock_etal:2019,Feeney:2020kxk}. While no reweighting is necessary in this case, the dependence on gridded computations means this method scales very poorly with parameter-set dimensionality.

Recently, Ref.~\cite{Talbot:2020oeu} proposed a machine-learning based approach to this problem. The authors use a Gaussian mixture model to fit the distribution of object parameters found using a set of detected mergers drawn from a fiducial population. By dividing out the prior on the object parameters for the fiducial population, they obtain an estimate of the probability of detecting a merger given its parameters. This estimate can be combined with the prior on the object parameters for a generic population to calculate $\Nbar$ at any point sampled, either directly or via a neural-network-based interpolation. This approach suffers less bias than the reweighting method due to the assumption of a fiducial population, and comes at a cost of only $O(1000)$ simulated populations. However, the estimate of the detection probability as a function of object parameters is only defined over the range of parameters supported by the fiducial population; should this range change rapidly with the population parameters, the method's $\Nbar$ estimates will lose accuracy.

Here, we take a different approach, demonstrating that the computation of $\Nbar$ can be completely bypassed using Likelihood-Free Inference (LFI), which requires no analytic knowledge of the likelihood function. Specifically, we use Density-Estimation LFI (DELFI) \cite{LFI_Papamakarios, LFI_Lueckmann,LFI,pydelfi}, in which the distribution of data as a function of the parameters that generated them is fit by supplying density estimators with a training set of simulated datasets. This fit is then used as a proxy likelihood to obtain posteriors on the parameters of interest. As the simulated data include the selection function, LFI automatically accounts for the Malmquist bias.

LFI's ability to accelerate the inference of the properties of individual BBH mergers has been demonstrated in a number of recent works~\cite{George_etal:2018, Shen_etal2019, Gabbard_etal:2019, Chua_etal:2020, Green_etal:2020, Green:2020dnx, Delaunoy_etal:2020}. Here, we apply LFI to population-level inference, taking as our example the inference of $\hubble$ from 100 simulated GW-selected BNS mergers with EM counterparts. In this particular setting, traditional Bayesian inference (with $\Nbar$ interpolated from a grid of cosmological values~\cite{Mortlock_etal:2019}) is feasible, and we take this approach as a ground truth from which we can robustly quantify any systematic errors introduced by LFI. We take as our inputs sets of individual mergers' observed redshifts, distances (generated via traditional~\cite{TheLIGOScientific:2017qsa} or likelihood-free analyses~\cite[e.g.][]{Green:2020dnx}) and peculiar velocities, performing our LFI analysis with the aid of \texttt{pydelfi} \cite{pydelfi}. While we concentrate here on the inference of $\hubble$ from BNS, the technique is applicable to population studies in general \cite[e.g.][]{Abbott:2020gyp, Kim_2021}. 

We describe the hierarchical model we use to simulate our BNS mergers in Sect.~\ref{sect:simulations}, and explain our inference method in Sect.~\ref{sect:method}, highlighting the importance of data compression. Results are discussed in Sect.~\ref{sect:results}, and conclusions are drawn in Sect.~\ref{sect:conclusions}.

\section{SIMULATIONS}\label{sect:simulations}

\begin{figure}
    \includegraphics[width=8.5cm]{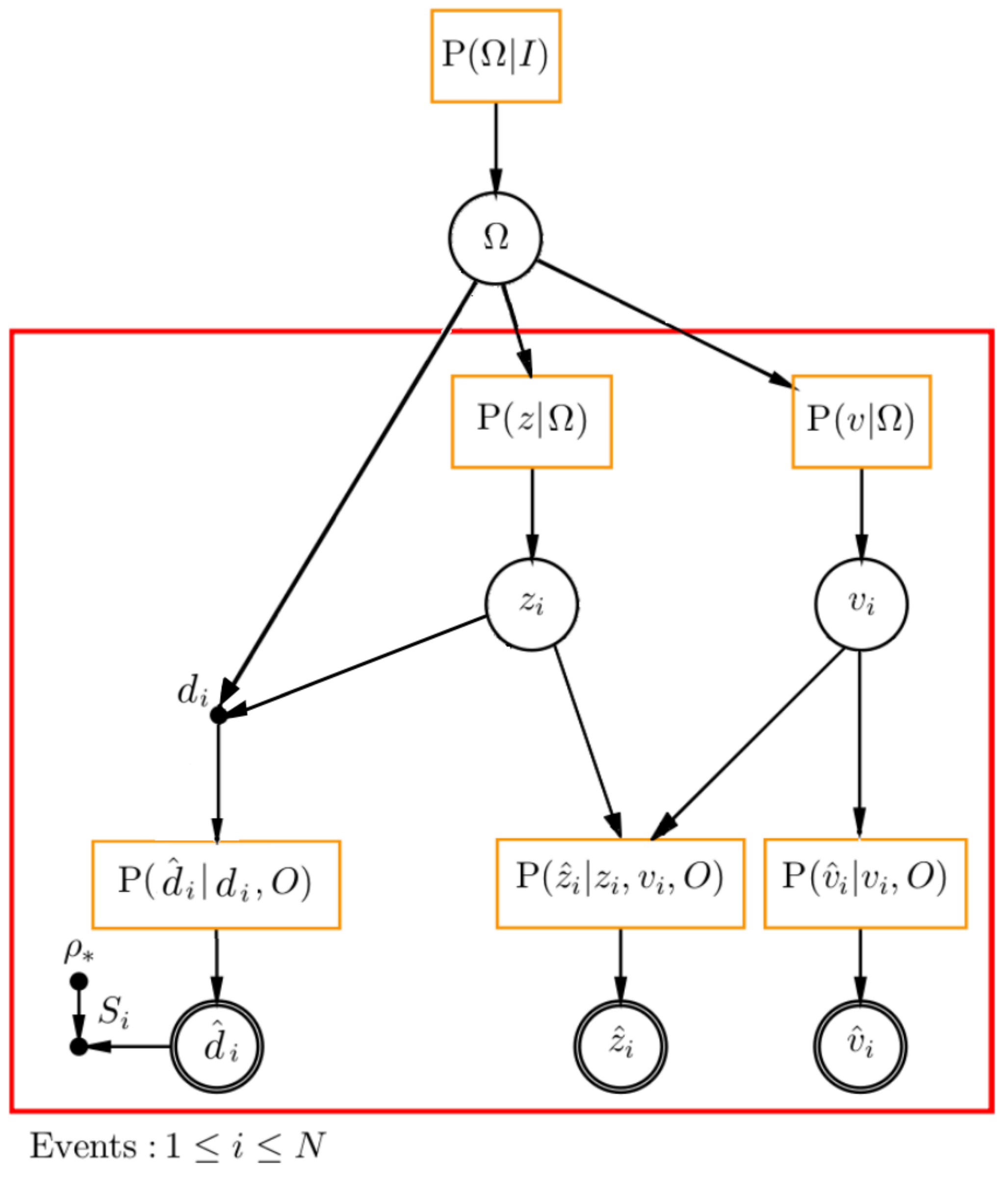}
    \caption{The hierarchical model used to describe our BNS population and data, adapted from Ref.~\cite{Mortlock_etal:2019}. Read top-to-bottom, parameters (circles) are drawn from probability distributions (orange rectangles) to generate observed quantities (double circles). \textit{I} represents the prior information assumed about the cosmological parameters, $\cosmo=[H_0,q_0]$, and quantities within the red plate are specific to an individual merger.}
    \label{fig:sims_diagram}
\end{figure}

In this work we assume we possess noisy estimates of redshift $\hz$, distance $\hd$ and peculiar velocity $\hv$ for each BNS merger. The mergers' $[\obsz,\obsd,\obsv]$ are generated via the hierarchical model in Fig.~\ref{fig:sims_diagram}, which is loosely based on the model used in Ref.~\cite{Mortlock_etal:2019}. We assume that the strain data have been pre-compressed into estimates of $\obsd$, which can be done rapidly using the likelihood-free method of Ref.~\cite{Green:2020dnx}. Given the aforementioned prospects for solving the $\hubble$ tension, we fix the number of mergers to $N=100$. We consider two test cases, both assuming the same set of observables, but distinguished by whether GW selection is applied. Considering these two cases allows us to differentiate the impact of LFI alone from LFI specifically in the presence of selection effects.

In the following we wish to infer two cosmological parameters -- the Hubble constant, $\hubble$, and the deceleration parameter, $\dec$ -- which we denote by $\cosmo=[\hubble,\dec]$.
For a given choice of $\cosmo$, true redshifts are randomly sampled from
\begin{eqnarray}\label{eqn:truered_probs}
    &P&(\zi | \cosmo,\zmax) \nonumber\\ 
    &&= \dfrac{1}{(1+\zi)}\dfrac{dV}{dz}(\cosmo) \pH(\zmax - \zi)  \nonumber\\ 
    && \simeq \dfrac{4\pi}{(1+\zi)} \dfrac{c^3 z^2}{\hubble^3}[1-2(1+\dec)\zi] \pH(\zmax - \zi),
\end{eqnarray}
where $\pH$ is a Heaviside step function. The final line is a good approximation for $z_{max} \ll 1$. Given a single cosmological redshift draw, the $i^{\rm{th}}$ distance is given by \cite{Visser:2003vq}
\begin{equation}\label{eqn:truedist_probs}
    \di (\zi,\hubble,\dec)  = \dfrac{c\zi}{\hubble} \left[  1 + \dfrac{1}{2} (1-\dec)\zi \right].
\end{equation}
Denoting as $\pN(\mu,\sigma)$ the normal distribution of mean $\mu$ and standard deviation $\sigma$, peculiar velocities are sampled from
\begin{align}
    P(\vi) &= \pN(\mu_{v_{\parallel}},\sigma_{v_{\parallel}}) \nonumber\\
    &= \pN(0 \;\vunits,500 \;\vunits).
    \label{eqn:truevel_probs}
\end{align}
We convert our true redshifts, distances and peculiar velocities into observed quantities $\obsx=[\obsz,\obsd,\obsv]$ assuming Gaussian noise as follows
\begin{eqnarray}
    P(\hz|z,v) & = & \pN (z+v/c, \sigma_{\hz} = 1.2 \times 10^{-3}) \\
    P(\hd|d) & = & \pN (d, \sigma_{\hd} = d/10)\\
    P(\hv|v) & = & \pN (v, \sigma_{\hv} = 200 \;\vunits).
\end{eqnarray}

\begin{figure}
    \includegraphics[width=8.5cm]{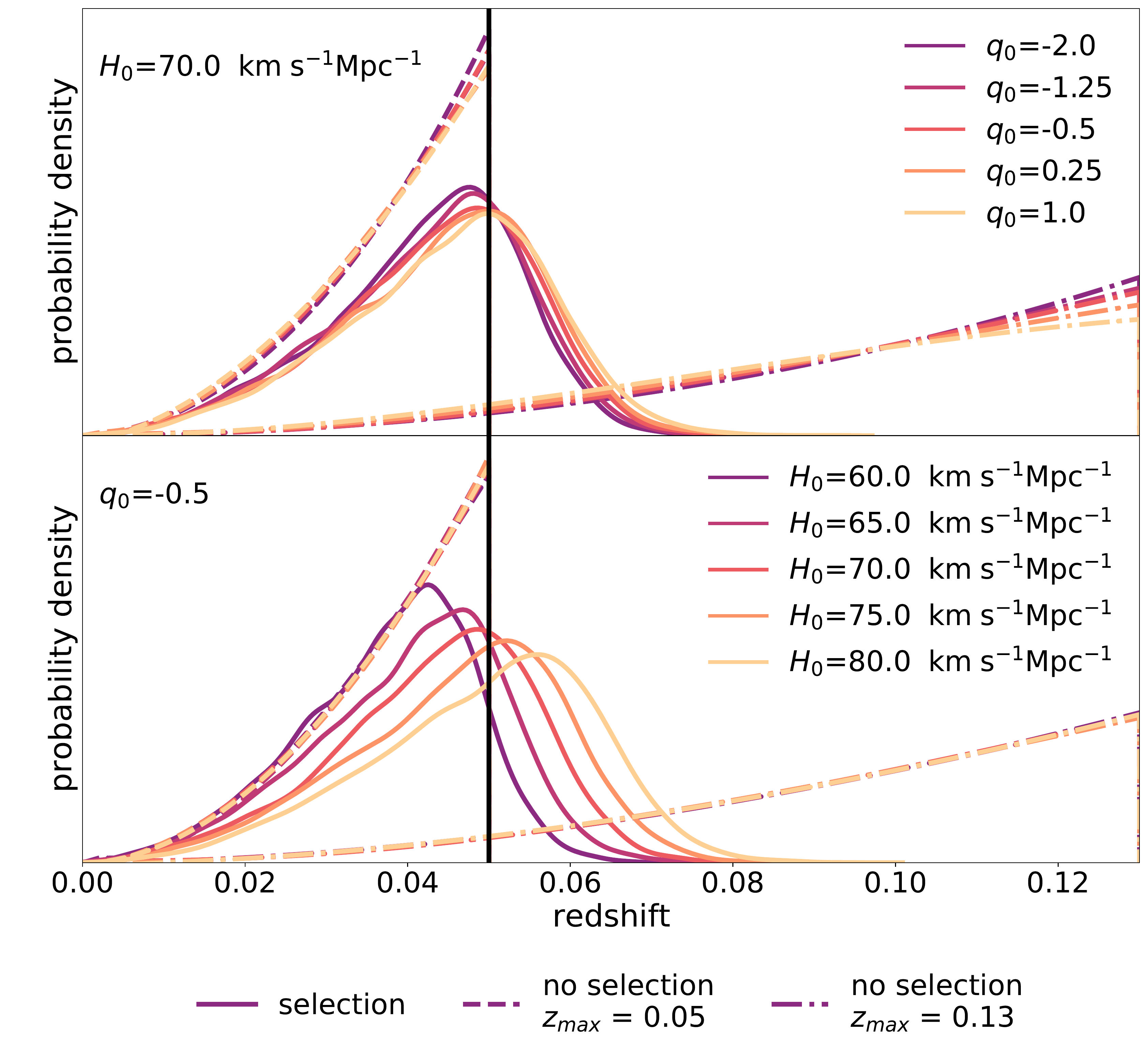}
    \caption{The dependence of BNS redshift distributions on $\dec$ (top) and $\hubble$ (bottom) for our no-selection (dashed) and selection datasets (solid). To obtain comparable constraints on $\hubble$ from the two datasets, we impose a cutoff at $z_{\rm max}=0.05$ for the no-selection case, while using $z_{\rm max}=0.13$ for the selection case. The input distribution for the selection dataset is shown as a dot-dashed line.}\label{fig:3D_redshifts}
\end{figure}

When GW selection is not applied, we simulate populations by simply drawing from the above distributions N times. When using GW selection, we require that the signal-to-noise ratio (SNR), defined as 
\begin{equation}
        \rho_i(\obsdi) = 12 \, \left( \dfrac{250\, {\rm Mpc}}{\obsdi} \right),
\end{equation}
is greater than $\thr =12$ for $i=[1,N]$. Introducing the GW selection changes the distribution of GW sources, reducing the effective upper redshift limit in a cosmology-dependent way, as shown in Fig.~\ref{fig:3D_redshifts}; the peak of the redshift distribution broadens and shifts to higher $z$ for increasing $\hubble$, while $\dec$ has a much smaller impact over this redshift range. For values of $\hubble \in [60,80]\;\hunits$ and $\dec \in [-2,1]$, the redshift distribution is peaked at $z \simeq 0.05$. To ensure we generate sources at similar redshifts for our selection and no-selection populations (and consequently obtain similar constraints on cosmological parameters) we set $\zmax$ equal to $0.05$ and $0.13$ for the no-selection and selection cases, respectively.

\section{METHOD}\label{sect:method}

\subsection{Traditional Inference}

We begin by outlining the traditional approach to inferring parameters from GW-selected populations, before describing our adopted likehood-free methodology. The traditional framework has been set out in numerous references~\cite{Schutz,Dalal:2006qt,Nissanke:2009kt,PhysRevD.85.023535,Nissanke:2013fka,Abbott_H0infer,Mandel:2018mve,Chen_BNSpredictions,Feeney_BNSpredictions,Gray:2019ksv,Vitale:2020aaz,Mortlock_etal:2019}, but we will follow the notation of Ref.~\cite{Mortlock_etal:2019} here. For simplicity, in this work we set aside the inference of the BNS properties (e.g. the NS mass distribution) and focus on the cosmology. As we are considering a fixed sample size here, the posterior on the cosmological parameters given a catalogue $\obsx=[\obsz,\obsd,\obsv]$ can be written as

\begin{align}
    P&(\boldsymbol{z},\boldsymbol{v},\hubble,\dec | \obsx) \propto \frac{P(\hubble)P(\dec)}{ \left[\Nbar(\hubble,\dec)\right]^N } \times \\
    & \prod_{i=1}^{N} P(\zi|\hubble,\dec,z_{\rm max}) P(\vi) P(\obszi|\zi,\vi) P(\obsdi|\di) P(\obsvi|\vi). \nonumber
\end{align}
We assume truncated Gaussian priors on the cosmological parameters
\begin{align} 
    \label{eqn:priors}
    P(\hubble) = & \pH(\hubble - 60)  \pH(80 - \hubble) \nonumber\\
    & \pN \left( 70\;\hunits, 20\;\hunits \right) \nonumber \\
    P(\dec) = & \pH(\dec + 2) \, \pH(1-\dec) \,  \pN \left( -0.55 , 0.5 \right).
\end{align}
All other distributions are taken to match those set out in Sect.~\ref{sect:simulations}.

The impact of the selection function is captured by the factor of  $\left[\Nbar(\hubble,\dec)\right]^{-N}$. $\Nbar$ (which, recall, denotes the expected number of \textit{detected} mergers) must be evaluated at every point in parameter space sampled by a particular inference tool. Here, we follow Ref.~\cite{Mortlock_etal:2019} in evaluating $\Nbar$ on a $10 \times 10$ grid in $\hubble$ and $\dec$ (boosting the fiducial detection rate $\Gamma = 1540 \rm{Gpc^{-3}yr^{-1}}$ \cite{TheLIGOScientific:2017qsa} by a factor of $130$ to reduce sample variance), and then fitting using a fourth-order (15-coefficient) polynomial.
%For each point of the grid we simulate 
%\begin{equation}\label{eqn:N_bar}
%    N = \int_{0}^{\zmax} %\dfrac{\Gamma}{(1+\zi)}\dfrac{dV}{dz}(\hubble,\dec)
%\end{equation}
%sources. 
Following Ref.~\cite{Mortlock_etal:2019}, we then perform traditional Bayesian Inference using No-U-Turn-Sampling \cite{2011arXiv1111.4246H} as implemented in the \texttt{pystan} package~\cite{stan,stan_2}, explicitly sampling each merger's true redshift and peculiar velocity along with $\hubble$ and $\dec$. We take the marginal posteriors on $\hubble$ and $\dec$ output by \texttt{pystan} as the ground truth in the tests that follow.

\subsection{Likelihood-Free Inference}

Explicitly calculating $\Nbar(\hubble,\dec)$ at each point of parameter space sampled is computationally unfeasible. The methods proposed to circumvent this issue must balance computational cost and accuracy. The standard method of estimating $\Nbar$ via a reweighted sum over a set of detected mergers generated using a fiducial population~\cite{Tiwari:2017ndi,2019RNAAS...3...66F,Abbott:2020gyp} works well provided the object-level parameter distribution for generic population parameters does not differ too strongly from that of the fiducial population~\cite{2019RNAAS...3...66F}. To counter this, the fiducial detected merger population must be oversampled, increasing the cost of both generating the detected sample and evaluating the likelihood. The cost of the former will become prohibitive in any setting where the distributions of object parameters have finite (or strongly suppressed) support which changes with the population parameters. Ref.~\cite{Talbot:2020oeu} estimates $\Nbar$ by fitting the distribution of object parameters found in the fiducial detection set and from this obtaining an estimate of the probability of detecting a merger given its parameters. This reduces both the computational cost and the bias due to estimating the detection probability from a fiducial population that might differ strongly from the underlying truth; however, it still fundamentally depends on the assumption of a fiducial population. The gridded approximation~\cite{Mortlock_etal:2019} we use for our traditional Bayesian analysis here does not require a fiducial population but is computationally expensive, requiring $\sim 130 \times N$ selected mergers for each single point of the grid, hence $\sim 13000$ detected samples in total. It can not be scaled to problems with a large number of population parameters.

Here we demonstrate that we can bypass the $\Nbar$ calculation entirely using likelihood-free methods, which are based solely on simulations and therefore naturally account for selection effects. In particular, we use Density-Estimation Likelihood Free Inference (DELFI) \cite{LFI_Papamakarios, LFI_Lueckmann,LFI, pydelfi}, in which synthetic mergers sampling the joint parameter-data space $(\cosmo,\obsx)$ are used to train neural density estimators (NDEs) to fit $P(\obsx|\cosmo)$, the probability of obtaining GW-selected data given the population parameters. By fitting this distribution, we implicitly marginalize over the mergers' true redshifts and peculiar velocities. The fit is evaluated at the observed data $\obsx_{\rm{obs}}$ to obtain $P(\obsx_{\rm obs}|\cosmo;\mathbf{w})$, a parametric model for the likelihood depending on the trained weights $\mathbf{w}$ of the neural density estimators. This is then multiplied by the prior to yield the final posterior $P(\cosmo|\obsx_{\rm obs}) \propto P(\cosmo)P(\obsx_{\rm obs}|\cosmo;\mathbf{w})$.
%$P(\cosmo|\obsx,\thr) \propto P(\cosmo)P(\obsx|\cosmo,\thr)$.

Our LFI analysis uses \texttt{pydelfi}\footnote{\url{https://github.com/justinalsing/pydelfi}}, an implementation of DELFI developed by Ref.~\cite{pydelfi}, based on Refs~\cite{LFI_Papamakarios,LFI_Lueckmann,LFI}. \texttt{pydelfi} learns a parametric model to the conditional distribution $P(\obsx|\cosmo)$ -- via \textit{on-the-fly} or precomputed simulations -- using a set of NDEs. The NDE components can be freely chosen as a combination of mixture density networks (MDNs) and masked autoregressive flows (MAFs) (see Refs.~\cite{pydelfi,Bishop94mixturedensity,MAFS,MAFS2} for details on the NDEs). To reduce the possibility of pathological behavior from one particular NDE affecting our results, we create an ensemble of estimators by stacking together five MDNs (with one to five Gaussian components) and one MAF. We use the same ensemble of NDEs for all \texttt{pydelfi} runs. To reduce variance in our results, we train all of the NDEs using a fixed set of $2000$ simulated training populations, rather than letting the algorithm generate on-the-fly simulations. These training samples are obtained by uniformly drawing from $\hubble \in [60,80]\;\hunits$ and $\dec \in [-2,1]$. The choice of the training-set size is empirically driven by the estimators' efficiency: there exists a (setting-specific) limiting training-set size beyond which there is no significant improvement in the training \cite{pydelfi}. Reducing the training set to 1000 populations significantly impacts the quality of our results; boosting it to 10000 does not improve the results enough to justify the higher computational cost.

\subsubsection{Data compression method}

As the simulated catalogues consist of $N = 100$ sources, performing LFI on the raw data would require fitting a $302$-dimensional probability distribution, which is unfeasible (given the available resources in terms of number of simulations and our fidelity requirements). In order to reduce the dimensionality of the inference space, the data must be compressed to a set of summary statistics $\obst$, a vector of $\mathrm{dim}(\obst)\equiv \mathrm{dim}(\cosmo)$ components (i.e., one compressed summary per parameter of interest). Identifying suitable summary statistics translates into finding a map $\mathit{f}: \obsx \rightarrow \obst$ that compresses the data while retaining as much information as possible. Methods capable of performing such a mapping include score compression \cite{Alsing:2017var,LFI,Alsing:2019dvb}, Information Maximizing Neural Networks \cite{2018PhRvD..97h3004C} and regression neural networks (NNs) \cite{10.5555/1162264}. In this work, we train regression neural networks to compress generic merger data into estimates of the generative cosmological parameters. For training purposes, we need to construct a set of training and validation datasets, for which the underlying cosmology is known and will constitute the target. The network will ultimately compress the noisy data to a set of summary statistics which correspond to a prediction about the generative cosmological model. To avoid any dependence on the particular training initialization of a single network, we create an ensemble of 9 trained neural networks, all defined by the same settings and trained on the same exact data but using different random initial weights.

The raw observables span a broad range of magnitudes -- $\hz \simeq O(10^{-2})$, $\hd \simeq O(10^{2})$ and $\hv \simeq O(10^{3})$ -- which can cause problems in the training process. If there are large differences in scale between different components of the data vector, the NN will naturally prioritize the larger components, effectively ignoring part of the dataset. Moreover, the magnitude of the data vector determines the update rate, so large values might lead to stability problems. Prior to feeding data into any neural network, therefore, we normalize the data to ensure they are all at roughly the same scale. We first sort all merger catalogues by redshift to reduce the variability to which each NN input node is exposed. We then concatenate each catalogue's $\obsz$, $\obsd$ and $\obsv$ to create a single 300-element raw-input vector. Finally we shift and scale by the mean and standard deviation of 100 catalogues generated at our fiducial cosmology $[\hubble,\dec]=[70,-0.5]$ to create the normalized inputs for our regression networks. We also normalize the target parameters which generated the training and validation datasets, shifting and scaling their distributions to be within 0 and 1. The NN predictions -- our summary statistics -- are hence normalized estimates of the cosmological parameters.

\subsubsection{Data compression optimization}

The choice of architecture and settings for our neural networks is completely free, which poses an intimidating optimization problem over the vast number of possible NN architectures and settings. To define a NN we must choose an architecture, its activation function and training, by tuning batch size, learning rate and potentially employing regularization methods. We cannot reasonably explore all of these choices, and we therefore consider neural networks composed of two hidden layers, each made of 128 hidden units, fix the activation function to be a Leaky Relu~\cite{Maas13rectifiernonlinearities} with \texttt{alpha}$ = 0.01$,\footnote{\url{https://keras.io/api/layers/activation_layers/leaky_relu/}} and focus on finding the best combination of batch size $\nb$ and learning rate $\alpha$ from a small set of choices, namely $\nb = [100,500]$ and $\alpha = [10^{-4},5 \times 10^{-4},10^{-3}]$. To avoid potential overfitting, we consider regularization terms, which control the training while acting on the loss function, set to be the mean squared error (MSE). We toggle between Ridge and Lasso regression methods, which use L2 and L1 regularizations respectively \cite{Hastie}, and explore a few values of the parameters weighting the regularization term, $\lambda_{1,2}$, namely $\{  \lambda_{1,2}=0 \}$,$\{ \lo=0,\lt=[10^{-4},2 \times 10^{-4}] \}$ and $\{ \lo=[10^{-4},2 \times 10^{-4}],\lt=0 \}$. 
We define the optimal compressor as the NN for which \texttt{pydelfi} most faithfully reproduces \texttt{pystan}'s results for a range of $[\hubble,\dec]$. The process by which we determine the optimal NN settings is described in the following.

For each combination of batch size, learning rate and regularization, we first train the regression NN on a set of $\nt$ samples of known cosmology, validating with a further $\nv$ datasets. To determine the impact of the amount of training data available on the final inference, we consider two training set sizes, the first with $[\nt,\nv]=[5000,2000]$ and the second with $[500000,100000]$. In all cases, the generative cosmologies are sampled from $\hubble \in [60,80]\;\hunits$ and $\dec \in [-2,1]$ using the Latin hypercube method. 

To determine the NN parameters that optimize LFI performance for a range of underlying cosmologies, we generate 100 test catalogues for cosmological parameters sampled from $\hubble \in [65,75]\;\hunits$ and $\dec \in [-0.7,-0.3]$ using the Latin hypercube method (the reason for this restricted range will be explained in Sect.~\ref{sect:results}). We then perform traditional Bayesian inference and LFI on each test catalogue, for each choice of NN parameters. Given these results, we compute the differences $\biasH = \tradH - \LFIH$ and $\biasq =\tradq - \LFIq$ between the maximum-posterior estimates of the cosmological parameters from the traditional and LFI approaches. Compiling the results from all of the test catalogues, we calculate the means ($\bar{b}_{H_0,q_0}$) and standard deviations ($\sigma_{{b}_{H_0,q_0}}$) of the biases injected by LFI for each compression NN. The optimal compression network is chosen to be that which minimizes the standard deviation of the $\hubble$ bias, provided its mean bias is consistent with zero.

In addition to requiring LFI produces unbiased estimates of the cosmological parameters, we also want to ensure our compression is as lossless as possible, i.e., that the LFI and traditional constraints have similar $\hubble$ uncertainties. To do so, we need the total uncertainty in the LFI parameter constraints, which we approximate as the quadrature sum of the ``raw'' uncertainty of the LFI posteriors and the additional uncertainty due to the bias.\footnote{This is equivalent to marginalizing over an unknown additive bias, assuming the parameters and bias are independent and Gaussian-distributed.} We estimate the former by calculating the mean variance of the LFI cosmological parameter posteriors over all 100 test catalogues; the uncertainty on the bias is simply $\sigma_{b_{\hubble}}$. Hence, the increase in the $\hubble$ uncertainty expected from replacing traditional Bayesian inference with LFI in this setting can be estimated by calculating
\begin{equation}
\incr = 100 \times \left(  \dfrac{\sqrt{ \left(\sigma_{\rm LFI}^{\hubble} \right)^2 + \sigma^{2}_{\biasH}}}{\sigma_{\rm trad}^{\hubble}} -1 \right).
\end{equation}

\section{RESULTS}\label{sect:results}

We first consider the no-selection case to demonstrate the feasibility of LFI in this setting and obtain a baseline for its impact on the precision and accuracy of the inference. We then add in GW selection to determine whether selection specifically affects LFI's performance, and to provide a final estimate of the systematics.

\subsection{No-Selection Case}

\begin{figure}[tp]
    \centering
    \includegraphics[scale=0.25]{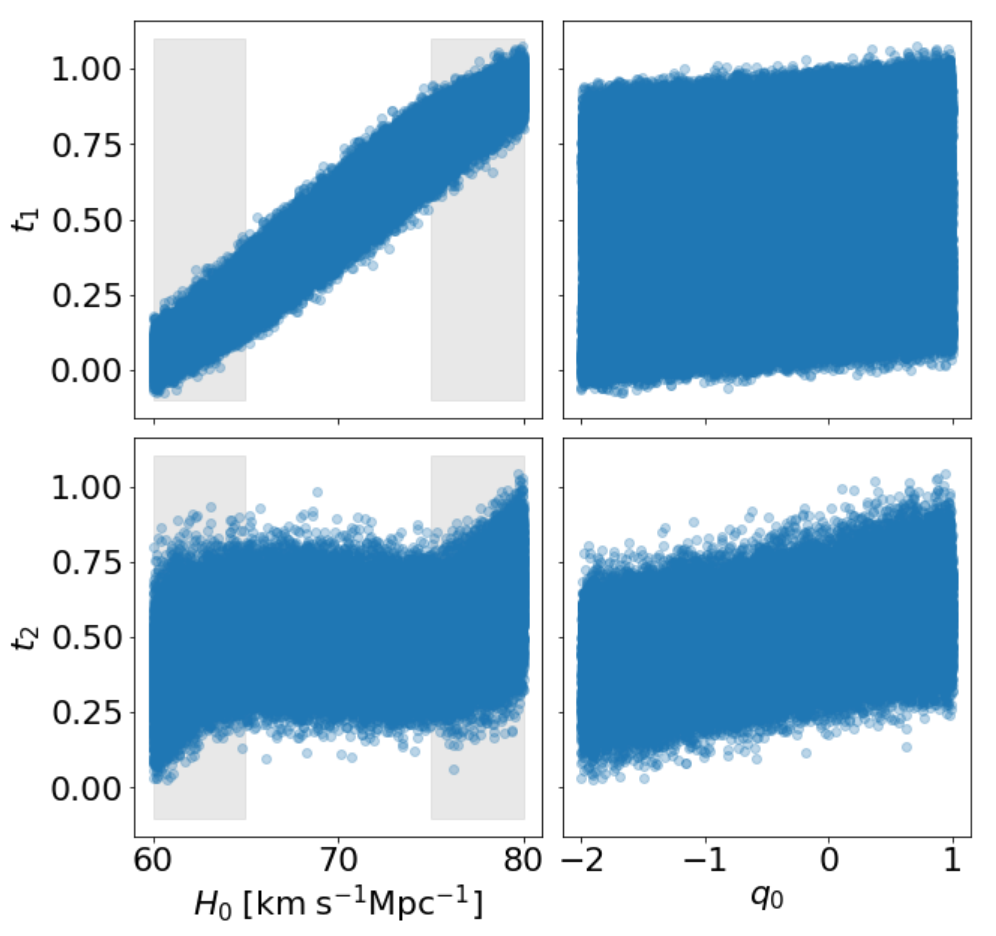}
    \caption{The summary statistics $\obst=(\hat{t}_1,\hat{t}_2)$ output by our compression NN plotted against the cosmological parameters at which the corresponding data were generated. This NN was trained with  $[\nb,\alpha,\lambda_{1,2}]=[100,10^{-4},0]$, and the points correspond to the validation dataset for the $[\nt,\nv]=[500000,200000]$ setup. The shaded areas indicate the regions of $\hubble$ where the slopes of the summary statistics change with respect to the central trend.} 
    \label{fig:3D_ssplot}
\end{figure}

Considering the no-selection case first gives us a baseline for gauging LFI's performance in the more complex setting with selection, allowing us to determine whether selection specifically has any impact on LFI. We train our compression NNs for all combinations of the aforementioned batchsize, learning rate and regularizer choices, for both training-set sizes $[\nt,\nv]$. Each of these neural networks provides different compression performance and thus all are tested as compressors in the LFI workflow. An example of compression performance for $[\nt,\nv]=[500000,200000]$ is given in Fig.~\ref{fig:3D_ssplot}, which shows the summary statistics $\obst$ output by the regression NN against the generative cosmological parameters for the validation set. Focusing on the $\hat{t}_1-\hubble$ and $\hat{t}_2-\hubble$ plots for now, we notice that the width and slope of the distribution change at the edges of the training set, shaded in grey. As the NN behaviour might be suboptimal in these ranges, we generate the test samples used to optimize the compressor settings from values of $\hubble$ within $[65,75]\;\hunits$, lying in the unshaded area.

\begin{figure*}[htp]
    \centering
    \includegraphics[scale=0.3]{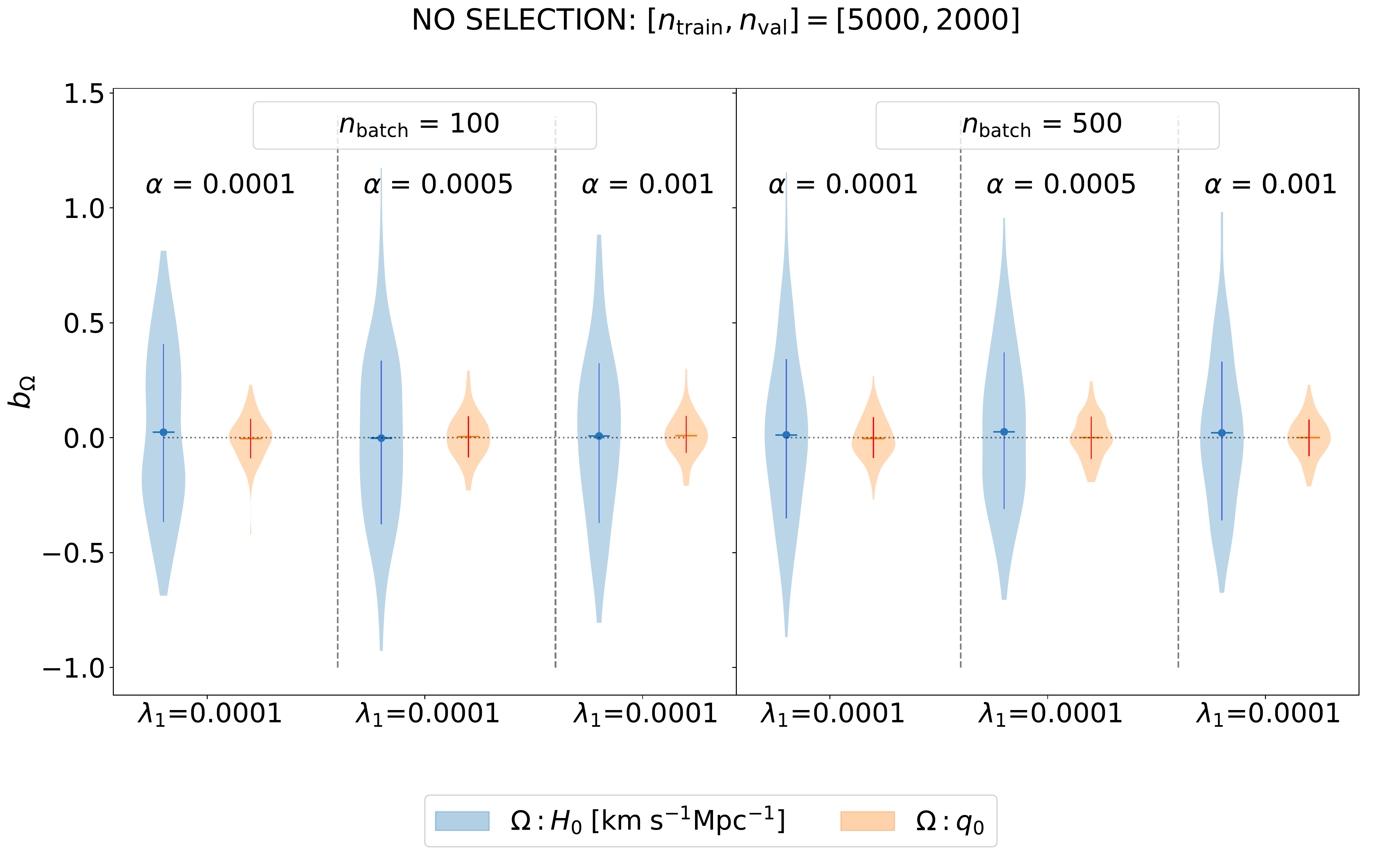}
    \caption{Violin plots for the $\biasH = \tradH - \LFIH$ (blue) and $\biasq = \tradq - \LFIq$ (orange) bias distributions for the no-selection setting. Results are shown for the NNs whose regularization choice minimizes the bias for each combination of batchsize $\nb$ and learning rate $\alpha$. Dots represent the mean biases, and lines the $1\sigma$ errorbars. The mean biases are consistent with zero, and the bias distributions are considerably narrower than the relevant parameter posteriors, for all NNs plotted.} 
    \label{fig:violin_3Dnosel_sum}
\end{figure*}

\begin{table*}[tp]
\caption{\label{tab:3D_nosel_results_sum} Means and standard deviations for the biases ${b}_{H_0,q_0}$, posterior-width ratios ${f}_{H_0,q_0}$ and percentage increase in $H_0$ uncertainty for the NNs whose regularization choice minimizes the bias for each combination of batchsize $\nb$ and learning rate $\alpha$ in the no-selection case.}
\begin{ruledtabular}
\begin{tabular}{@{} cccccccc}
\multicolumn{8}{c}{\textbf{NO SELECTION CASE}} \\ \hline \hline 
 $\nb$ & $\alpha$ & regularizer & $\biasH \;[\hunits]$ & $\biasq$ & $\fsH$ & $\fsq$ & $\incr $ \\
  \hline \hline
  \rowcolor{Gray}
 \multicolumn{8}{c}{\textbf{TRAINING and VALIDATION parameters:} $[\nt,\nv]=[5000,2000]$} \\ \hline \hline
 \multirow{3}{*}{$100$} & $10^{-4}$ & $\lo = 10^{-4}$  & $0.024 \pm 0.35$ & $-0.003 \pm 0.095$ & $1.014 \pm 0.045$ & $0.95 \pm 0.035$ & $7.64\%$\\
    & $5 \times 10^{-4}$ & $\lo = 10^{-4}$  & $-0.002 \pm 0.365$ & $0.004 \pm 0.098$ & $1.028 \pm 0.042$ & $0.952 \pm 0.032$ & $9.43\%$\\
    & $10^{-3}$ & $\lo = 10^{-4}$  & $0.007 \pm 0.352$ & $0.009 \pm 0.09$ & $1.024 \pm 0.048$ & $0.952 \pm 0.038$ & $8.58\%$\\ \hline
   \multirow{3}{*}{$500$} & $10^{-4}$ & $\lo = 10^{-4}$  & $0.012 \pm 0.358$ & $-0.003 \pm 0.092$ & $1.003 \pm 0.043$ & $0.947 \pm 0.036$ & $6.81\%$\\
    & $5 \times 10^{-4}$ & $\lo = 10^{-4}$  & $0.026 \pm 0.328$ & $0.001 \pm 0.091$ & $1.018 \pm 0.051$ & $0.948 \pm 0.026$ & $7.3\%$\\
    & $10^{-3}$ & $\lo = 10^{-4}$  & $0.021 \pm 0.322$ & $-0.0 \pm 0.087$ & $1.012 \pm 0.054$ & $0.943 \pm 0.036$ & $6.45\%$\\
 \hline \hline 
 \rowcolor{Gray}
 \multicolumn{8}{c}{\textbf{TRAINING and VALIDATION parameters:} $[\nt,\nv]=[500000,100000]$} \\ \hline \hline 
\multirow{3}{*}{$100$} & $10^{-4}$ & $\lt = 2 \times 10^{-4}$  & $-0.073 \pm 0.193$ & $0.015 \pm 0.061$ & $0.979 \pm 0.042$ & $0.945 \pm 0.038$ & $-0.05\%$\\
    & $5 \times 10^{-4}$ & --  & $-0.061 \pm 0.218$ & $0.014 \pm 0.071$ & $0.978 \pm 0.048$ & $0.948 \pm 0.04$ & $0.35\%$\\
    & $10^{-3}$ & --  & $-0.058 \pm 0.21$ & $0.02 \pm 0.058$ & $0.973 \pm 0.042$ & $0.945 \pm 0.04$ & $-0.35\%$\\ \hline
   \multirow{3}{*}{$500$} & $10^{-4}$ & $\lt = 10^{-4}$  & $-0.043 \pm 0.193$ & $0.017 \pm 0.066$ & $0.972 \pm 0.041$ & $0.944 \pm 0.039$ & $-0.77\%$\\
    & $5 \times 10^{-4}$ & $\lt = 2 \times 10^{-4}$  & $-0.053 \pm 0.208$ & $0.015 \pm 0.061$ & $0.975 \pm 0.046$ & $0.944 \pm 0.037$ & $-0.15\%$\\
    & $10^{-3}$ & $\lt = 10^{-4}$  & $-0.062 \pm 0.189$ & $0.012 \pm 0.06$ & $0.979 \pm 0.048$ & $0.946 \pm 0.035$ & $-0.22\%$\\
\end{tabular}
\end{ruledtabular}
\end{table*}

We identify the best regularization for each combination of batchsize and learning rate using the $\biasH$ distribution. The $\biasH$ and $\biasq$ probability densities are respectively shown as blue and orange violin plots in Fig.~\ref{fig:violin_3Dnosel_sum}, for $[\nt,\nv]=[5000,2000]$, and summarized in Table~\ref{tab:3D_nosel_results_sum}. Results for all NN parameter choices can be found in Tables~\ref{tab:3D_nosel_results} and~\ref{tab:3D_nosel_results_large}. From the violin plots we see that the likelihood-free inference of both $\hubble$ and $\dec$ is unbiased, since the bias is consistent with zero for all choices of NN parameters. For the best models, independent of the specific NN parameters and data realization, LFI's maximum posterior estimate for both parameters is typically well within \texttt{pystan}'s $1\sigma$ posterior uncertainty ($\geq 0.89 \,\hunits$ for these test populations).

We observe that for our smaller training set, regularization greatly improves performance. As an example, considering $[\nb,\alpha]=[100,10^{-4}]$ we find that adding a regularization term $\lo = 10^{-4}$ reduces $\sbiasH$ from $1.75$ to $0.35$ and markedly increases the $\hubble$ constraining power, reducing $\fsH = \sigma_{\hubble}^{\rm LFI} / \sigma_{\hubble}^{\rm trad}$ from $1.95$ to $1.06$. With regularization added, the width of the LFI $\hubble$ posterior is compatible with \texttt{pystan}'s. Considering the larger training set reduces the impact of the regularizer and significantly reduces the $\hubble$ LFI posterior's uncertainty, which we find to be systematically $\sim2-3\%$ smaller than \texttt{pystan}'s: we suspect that this is due to slight overfitting by \texttt{pydelfi}. The LFI $\dec$ constraints are also $\sim5\%$ tighter than \texttt{pystan}'s, independent of the size of the training set.

For the $[\nt,\nv]=[5000,2000]$ setup, the network with $[\nb,\alpha,\lo] = [500,10^{-3},10^{-4}]$ imparts the smallest bias in the $\hubble$ posterior, with $\sigma_{{b}_{H_0}}=0.32$. The $\hubble$ bias shrinks further when using our larger training set, with $\sigma_{{b}_{H_0}} = 0.19$. As the bias is small and consistent with zero it could be ignored when doing population-level inference; here, however, we marginalize over it and find that it would impart a 6.45\% and -0.05\% increase in the quoted $\hubble$ uncertainty, respectively: well within any reasonable tolerance. We note here that this slight increase in uncertainty is entirely down to imperfect compression, since in tests \texttt{pydelfi} provides the same posteriors when rerunning on the same compressed data.

\begin{figure}
    \centering
    \includegraphics[scale=0.3]{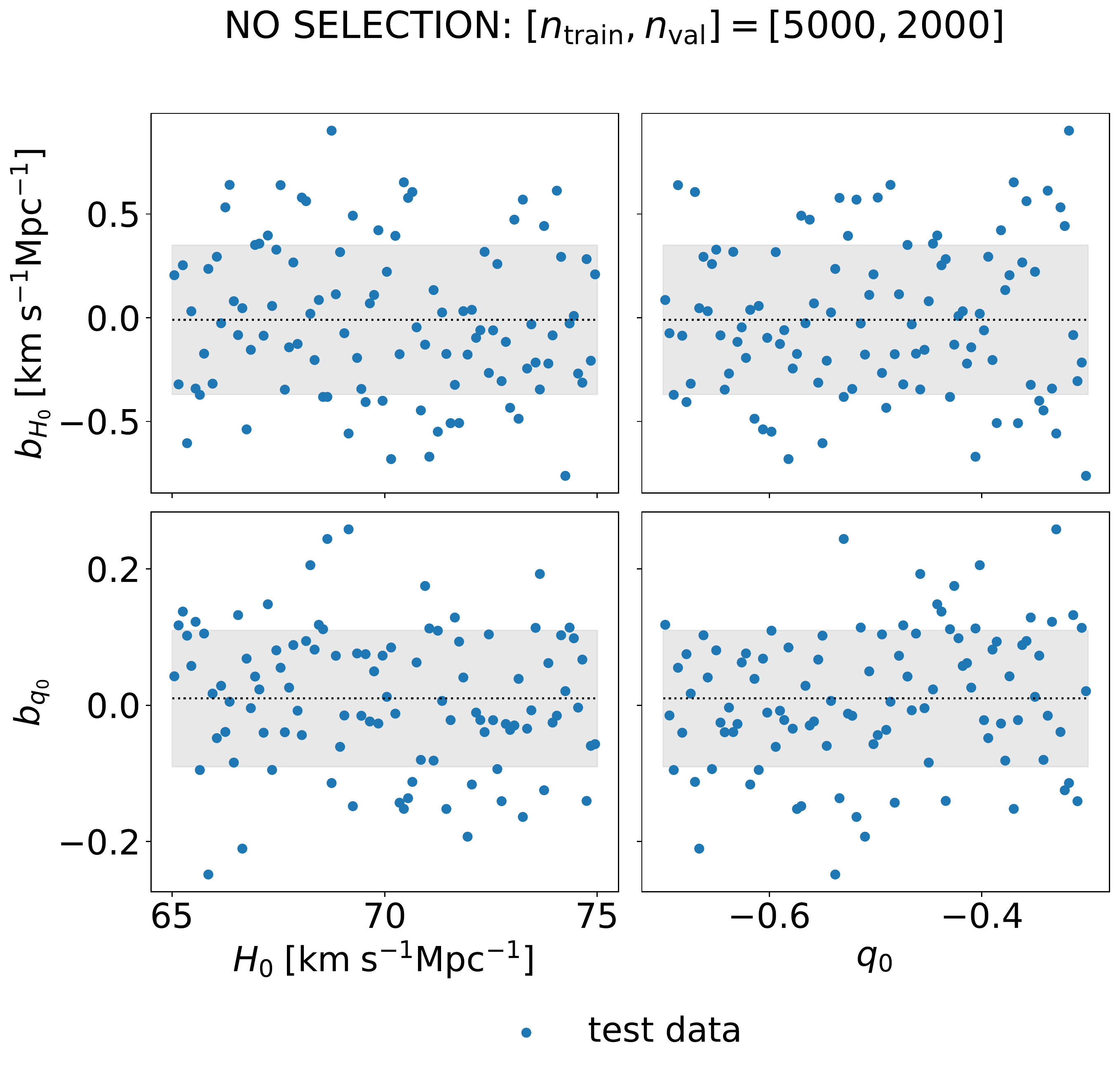}
    \caption{Distribution of generative parameters and LFI posterior biases. The one-sigma range of the bias is shaded grey. The neural network model used to perform the compression and generate this plot corresponds to the NN parameters combination $[\nb,\alpha,\lo] = [500,10^{-3},10^{-4}]$ for $[\nt, \nv] = [5000, 2000]$.} 
    \label{fig:3D_datapoints}
\end{figure}

One advantage of using a regression neural network for compression is that it only relies on a fiducial model for the computation of the mean and standard deviations used to normalize the neural network inputs. Nevertheless, the compression is sensitive to the choice of the training and validation data, as well as the range of sampled $\cosmo$ values. To investigate the randomness of the $\hubble$ bias with respect to the sampled parameter space, we plot the biases against the generative parameters for all 100 test catalogues for our best compression network in Fig.~\ref{fig:3D_datapoints}. We find there is no major correlation between the true parameters and the biases (for example, for the best model of the $[\nt, \nv] = [5000, 2000]$ setup, we find correlation coefficients of $C(\hubble,\biasH) = -0.13$ and $C(\dec,\biasH) = -0.023$).

\subsection{Selection Case}

\begin{figure*}[htp]
    \centering
    \includegraphics[scale=0.3]{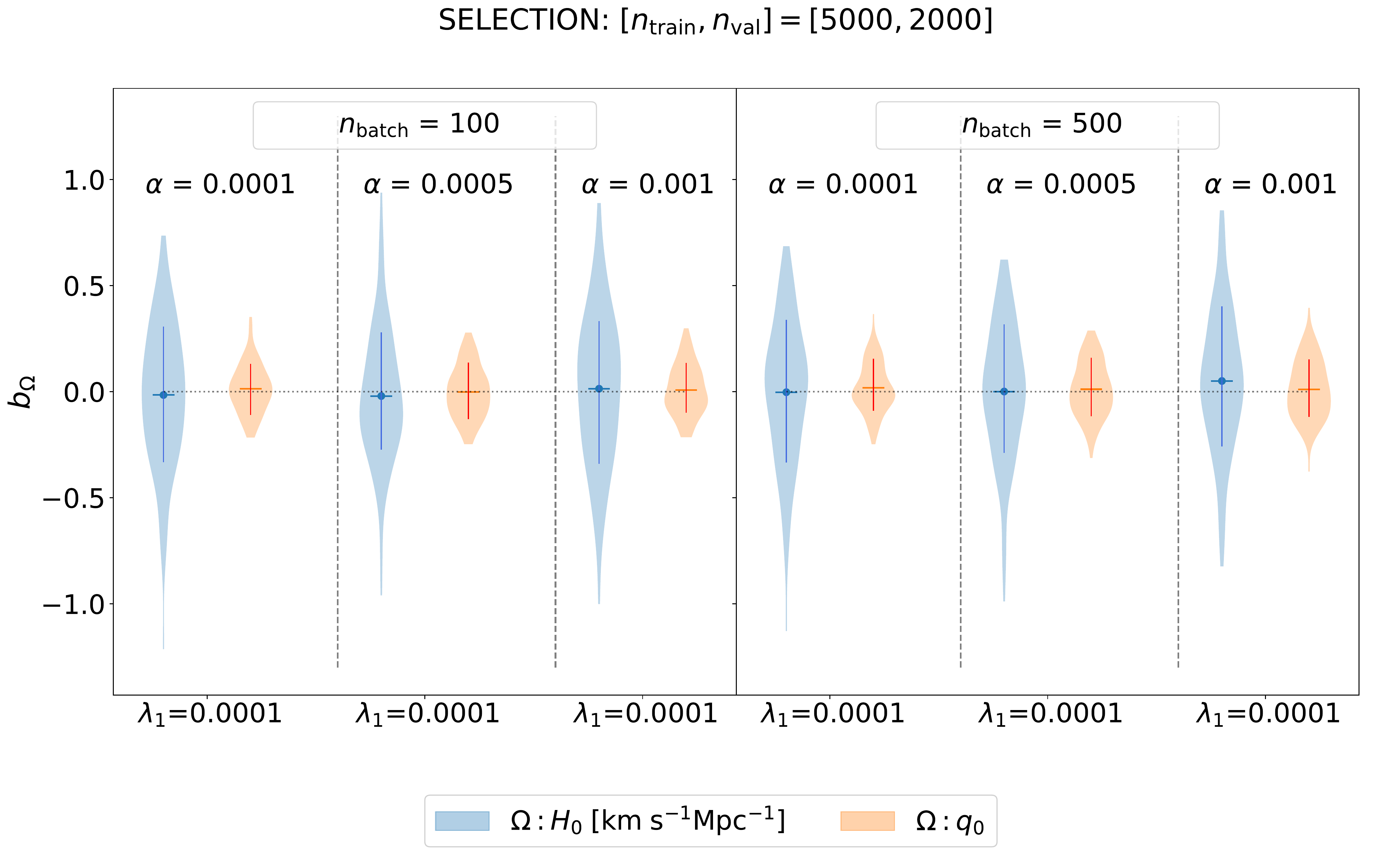}
    \caption{Violin plots for the $\biasH$ (blue) and $\biasq$ (orange) bias distributions for the setting with GW selection. Results are shown for the NNs whose regularization choice minimizes the bias for each combination of batchsize $\nb$ and learning rate $\alpha$. Dots represent the mean biases, and lines the $1\sigma$ errorbars. As in the no-selection case, the mean biases are all consistent with zero, and the bias distributions are all considerably narrower than the relevant parameter posteriors.} 
    \label{fig:violin_3Dsel_sum}
\end{figure*}

We now proceed to determine the impact of selection on the compression. As in the no-selection case, we first optimize the regularization for each combination of batchsize and learning rate. We compute the distributions of the $\hubble$ and $\dec$ biases, plotting the results for the best compressors in Fig.~\ref{fig:violin_3Dsel_sum} and tabulating their performance in Table~\ref{tab:3D_sel_results_sum}. Results for all the NN parameters can be found in Tables~\ref{tab:3D_sel_results} and~\ref{tab:3D_sel_results_large}. As in the no-selection case, the LFI maximum-posterior parameter estimates are unbiased when compared to the \texttt{pystan} baseline.

\begin{table*}[htp]
\caption{\label{tab:3D_sel_results_sum} Means and standard deviations for the biases ${b}_{H_0,q_0}$, posterior-width ratios ${f}_{H_0,q_0}$ and percentage increase in $H_0$ uncertainty for the NNs whose regularization choice minimizes the bias for each combination of batchsize $\nb$ and learning rate $\alpha$ in the selection case.}
\begin{ruledtabular}
\begin{tabular}{@{} cccccccc}
\multicolumn{8}{c}{\textbf{SELECTION CASE}} \\ \hline \hline \hspace{0.4cm}
 $\nb$ & $\alpha$ & regularizer & $\biasH \;[\hunits]$ & $\biasq$ & $\fsH $ & $\fsq$ & $\incr $ \\
  \hline \hline
  \rowcolor{Gray}
 \multicolumn{8}{c}{\textbf{TRAINING and VALIDATION parameters:} $[\nt,\nv]=[5000,2000]$} \\ \hline \hline
 \multirow{3}{*}{$100$} & $10^{-4}$ & $\lo = 10^{-4}$  & $-0.015 \pm 0.338$ & $0.014 \pm 0.115$ & $1.013 \pm 0.039$ & $1.005 \pm 0.044$ & $6.53\%$\\
    & $5 \times 10^{-4}$ & $\lo = 10^{-4}$  & $-0.02 \pm 0.313$ & $-0.001 \pm 0.122$ & $1.014 \pm 0.041$ & $1.005 \pm 0.058$ & $5.9\%$\\
    & $10^{-3}$ & $\lo = 10^{-4}$  & $0.014 \pm 0.357$ & $0.008 \pm 0.119$ & $1.018 \pm 0.042$ & $1.006 \pm 0.051$ & $7.67\%$\\ \hline
   \multirow{3}{*}{$500$} & $10^{-4}$ & $\lo = 10^{-4}$  & $-0.002 \pm 0.334$ & $0.019 \pm 0.116$ & $1.025 \pm 0.04$ & $1.01 \pm 0.049$ & $7.63\%$\\
    & $5 \times 10^{-4}$ & $\lo = 10^{-4}$  & $0.001 \pm 0.313$ & $0.012 \pm 0.128$ & $1.025 \pm 0.038$ & $1.013 \pm 0.036$ & $6.99\%$\\
    & $10^{-3}$ & $\lo = 10^{-4}$  & $0.051 \pm 0.329$ & $0.011 \pm 0.137$ & $1.019 \pm 0.053$ & $1.011 \pm 0.058$ & $6.91\%$\\

 \hline \hline 
 \rowcolor{Gray}
 \multicolumn{8}{c}{\textbf{TRAINING and VALIDATION parameters:} $[\nt,\nv]=[500000,100000]$} \\ \hline \hline 
  \multirow{3}{*}{$100$} & $10^{-4}$ & $\lt = 10^{-4}$  & $-0.032 \pm 0.184$ & $0.022 \pm 0.092$ & $0.976 \pm 0.031$ & $1.006 \pm 0.036$ & $-0.73\%$\\
    & $5 \times 10^{-4}$ & $\lt = 10^{-4}$  & $-0.033 \pm 0.177$ & $0.02 \pm 0.092$ & $0.979 \pm 0.039$ & $1.003 \pm 0.043$ & $-0.56\%$\\
    & $10^{-3}$ & --  & $0.0 \pm 0.183$ & $0.026 \pm 0.091$ & $0.965 \pm 0.03$ & $1.004 \pm 0.038$ & $-1.88\%$\\ \hline
   \multirow{3}{*}{$500$} & $10^{-4}$ & $\lo = 10^{-4}$  & $-0.022 \pm 0.178$ & $0.015 \pm 0.093$ & $0.978 \pm 0.036$ & $1.003 \pm 0.039$ & $-0.7\%$\\
   & $5 \times 10^{-4}$ & $\lt = 10^{-4}$  & $-0.013 \pm 0.18$ & $0.019 \pm 0.086$ & $0.977 \pm 0.035$ & $1.006 \pm 0.038$ & $-0.68\%$\\
    & $10^{-3}$ & $\lt = 2 \times 10^{-4}$  & $-0.01 \pm 0.199$ & $0.021 \pm 0.083$ & $0.979 \pm 0.043$ & $1.003 \pm 0.042$ & $-0.18\%$\\
\end{tabular}
\end{ruledtabular}
\end{table*}

As before, for our smaller training set regularization overall largely improves the performance. Considering $[\nb,\alpha]=[100,10^{-4}]$ as an example as before, we find that regularizing the training for $\lo = 10^{-4}$ reduces the uncertainty on the $\hubble$ bias from $1.71$ to $0.34$ and greatly improves the $\hubble$ constraining power, from $\fsH = 1.77$ to $1.06$. As in the no-selection case, the LFI posteriors produced using the optimal compressors are completely compatible with \texttt{pystan}'s. Again, increasing the training set size reduces the impact of the regularizer and significantly reduces the LFI $\hubble$ posterior's uncertainty, to $\sim 2.5\%$ smaller than \texttt{pystan}'s. 

For the $[\nt,\nv]=[5000,2000]$ setup two NN compressors minimize the $\hubble$ bias, with $\sigma_{{b}_{H_0}}=0.31$. These are defined by $[\nb,\alpha,\lo] = \{ [100,5 \times 10^{-4},10^{-4}],[500,5 \times 10^{-4},10^{-4}] \}$. As in the no-selection case the best models compressors use $\lo$ regularization. For the larger $[\nt,\nv]=[500000,100000]$ setup, the smallest standard deviation for the $\hubble$ bias is again considerably smaller: $\sigma_{{b}_{H_0}} = 0.18$ for the compressor with $[\nb,\alpha,\lt]=[100,5 \times 10^{-4},10^{-4}]$. As in the no-selection case, we compute the percentage increase in uncertainty on $\hubble$ imparted by replacing traditional inference with LFI, marginalizing over the bias. For the aforementioned three best compressors, these percentage increases are $\{ 5.9\%,6.99\% \}$ and $-0.56\%$, respectively, compatible with that determined for the no-selection case. Including GW selection does not impact LFI performance on a statistical level. Illustrative examples of the $\hubble$-$\dec$ joint posteriors produced by \texttt{pydelfi} and \texttt{pystan} can be found in Fig.~\ref{fig:posteriorplots}. 

\begin{figure}[htp]
    \centering
    \includegraphics[scale=0.32]{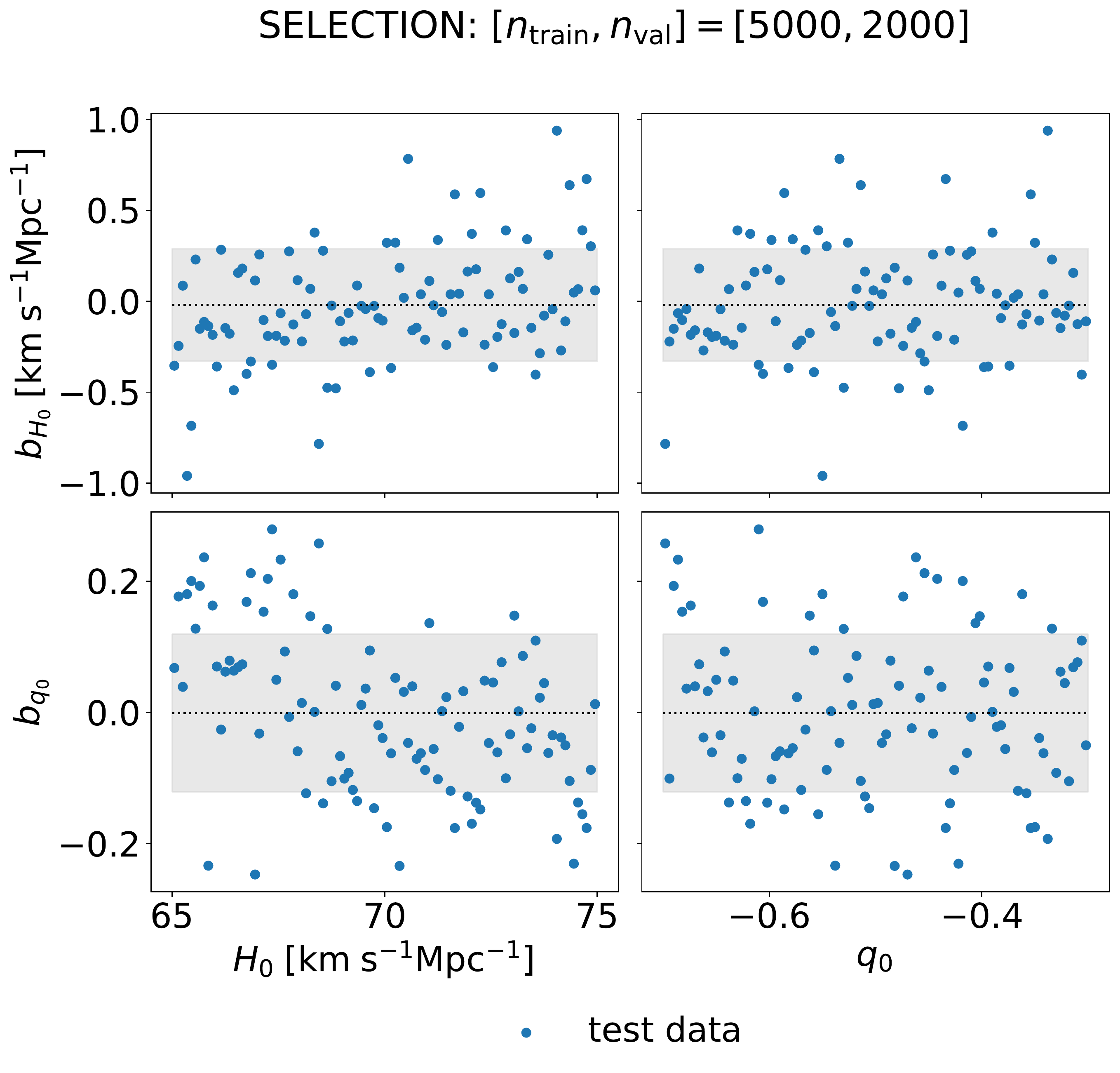}
    \caption{Distribution of generative parameters and LFI posterior biases for the GW selection setting. The one-sigma range of the bias is shaded grey. The neural network model used to perform the compression and generate this plot corresponds to the NN parameters combination $[\nb,\alpha,\lo] = [100,5 \times 10^{-4},10^{-4}]$ for $[\nt,\nv]=[5000,2000]$.} 
    \label{fig:3D_datapoints_sel}
\end{figure}

In Fig.~\ref{fig:3D_datapoints_sel} we plot the values of the $\biasH$ and $\biasq$ distributions against true input cosmology parameters. Unlike in Fig.~\ref{fig:3D_datapoints}, there is a clear dependence of $\biasH$ and $\biasq$ on the true value of $\hubble$ that generated the data. The strongest correlation is between the $\dec$ bias and the generative $\hubble$, with a correlation coefficient of $-0.47$ for the best model $[\nb,\alpha,\lo] = [100,5 \times 10^{-4},10^{-4}]$ of the smaller training set $[\nt,\nv]=[5000,2000]$. Increasing the size of the training sample generates stronger correlation ($-0.66$ for the best model). This indicates the regression is not capturing the selection function perfectly, and that other compression methods may fare better. Nevertheless, for the optimal compressors the biases on the cosmological parameters are consistent with zero, and have standard deviations which are a small fraction of the full posterior uncertainty.

\section{CONCLUSIONS}\label{sect:conclusions}

We have investigated the ability of Likelihood Free Inference (LFI) to estimate the cosmological expansion from GW-selected populations of binary neutron star mergers with EM counterparts. When computing the parameter posterior using traditional Bayesian inference, selection effects must be taken into account through the computation of the expected number of detected sources, $\Nbar$. This is a computationally expensive (and potentially inaccurate) process, even in approximate forms~\cite{Tiwari:2017ndi,2019RNAAS...3...66F,Mortlock_etal:2019}. As LFI does not explicitly evaluate the posterior, instead building a proxy likelihood using neural density estimator fits to parameter--simulated-dataset pairs, there is no need to calculate $\Nbar$ when performing LFI. Instead, the selection is naturally built into the simulations on which the method is based.

The goal of this work was to compare the precision and accuracy achievable using LFI to that of traditional Bayesian inference in the presence of selection effects. In this work we considered GW selection only; adding EM selection would increase the computational burden, making accounting for selection effects even more expensive. We employed ``pre-processed'' 100-merger datasets, consisting of noisy estimates of redshift, distance and peculiar velocity for each merger, assuming the distances have already been inferred from GW strains (which can be performed rapidly as in Ref.~\cite{Green:2020dnx} to yield a fully LFI-based pipeline). Given the high dimensionality of the input data, LFI methods require the data to be compressed to a set of summary statistics. We trained ensembles of regression neural networks for this purpose, passing their outputs to the density-estimation likelihood-free-inference package \texttt{pydelfi} to infer the cosmological parameters. Both of these stages require the provision of training data: we have presented results for compression networks trained using $[\nt,\nv]=[5000,2000]$ and $[\nt,\nv]= [500000,100000]$ populations; in all cases \texttt{pydelfi} was trained using 2000 simulated populations. Given each population contains 100 mergers, the total number of detected mergers required to train the two setups was $9 \times 10^5$ and $6 \times 10^8$, respectively.

LFI's precision and accuracy depends sensitively on the compression method's ability to retain salient information about the parameters of interest. We trained a large suite of regression networks (each containing two hidden layers of 128 hidden units) for compression, optimizing the learning rate, batch size and regularization based on \texttt{pydelfi}'s ability to infer $\hubble$ using the networks' outputs. Specifically, we selected the network whose resulting $\hubble$ inference best reproduced the traditional Bayesian ``ground truth'' (as implemented using \texttt{pystan}) for a set of 100 test datasets, taking the differences between maximum-posterior $\hubble$ estimates for the two methods as our metric.

Testing the method first on datasets in which no GW selection was made, we demonstrated that LFI provides unbiased $\hubble$ estimates when using suitably optimized regression-network data compression. For our optimal combination of training variables, we found a bias on $\hubble$ of $b_{H_0} = 0.021 \pm 0.322 \,\hunits$: consistent with zero and with a standard deviation a factor of roughly three smaller than the posterior uncertainty on $\hubble$. Marginalizing over this bias would lead to an increase of only $6.45\%$ in the uncertainty on $\hubble$. Adding in GW selection, we find no impact on LFI's performance: LFI is still able to provide unbiased estimates of $\hubble$ in the presence of selection effects. For the best model we obtain  $\biasH=-0.02\pm0.313\,\hunits$, which would yield an increase in uncertainty on $\hubble$ of only $5.9\%$ when marginalized over. Increasing the number of samples used to train the compression networks results in LFI posteriors that are statistically indistinguishable from their traditional Bayesian counterparts in mean and variance; however, this comes with a significant increase in computational cost. When processing GW-selected data, we note a small but significant correlation between the $\hubble$ and $\dec$ biases and the generative $\hubble$ values. This indicates a different choice of compressor architecture and setup might improve results, but investigating alternative compression methods is left for future work.

As this method is simulation-based, having a trustworthy and sufficient generative model is critical. This analysis has been conducted on simplified mock data, for which we know the underlying model. In the context of real observations, more-realistic simulations, such as those implemented in LALSuite \cite{LALSuite}, are needed. As current ground-based interferometers enhance their sensitivity~\cite{prospects}, third-generation GW detectors such as Einstein Telescope \cite{EinsteinTelescope} and Cosmic Explorer \cite{CosmicExplorer} come online, and the BNS sample builds, including instrumental systematics~\cite{Sun_etal:2020} and an as-yet elusive model of joint EM-GW selection~\cite[e.g.][]{Rosswog_etal:2017,Scolnic_etal:2018,Cowperthwaite_etal:2019,Setzer_etal:2019,Chen:2020,Mastrogiovanni_etal:2020,Feeney:2020kxk,Raaijmakers_etal:2021} will become ever more important. In this work we have focused on inferring the cosmological parameters only, but complete inference of the population properties of BNS catalogues must include parameters fixed here, such as the merger rate, mass distributions and equation of state~\cite[e.g.][]{LVC_NS_EOS:2018,Farrow_etal:2019,Landry_etal:2020,Abbott:2020gyp,Gauladage_etal:2021,Mastrogiovanni_etal:2021}. Extending the analysis to incorporate these parameters is left to future work. Finally, we note that, though we have focused on the inference of the cosmological expansion from GW-selected catalogues of binary neutron star mergers with EM counterparts here, this method can be applied to a broad range of population analyses in the presence of selection effects~\cite[e.g.][]{Abbott:2020gyp, Kim_2021}.

The code is provided at \url{https://github.com/frgerardi/LFIH0_BNS.git}.

\begin{acknowledgments}
We thank Tom Charnock, Will Farr and Colm Talbot for helpful discussions. This work was partially enabled by funding from the UCL Cosmoparticle Initiative. S.M.F. is supported by the Royal Society. J.A. was supported by the research project grant \emph{Fundamental Physics from Cosmological Surveys} funded by the Swedish Research Council (VR) under Dnr 2017-04212.

\end{acknowledgments}

%\nocite{*}

\bibliography{biblio}% Produces the bibliography via BibTeX.

\appendix

\counterwithin{figure}{section}
\counterwithin{table}{section}

\section{Full tables}

For completeness, in the following we tabulate the results for all combinations of learning rate, batchsize and regularization explored for both no-selection and selection analyses. 

\begin{table*}[tp]
\caption{\label{tab:3D_nosel_results} Means and standard deviations for the biases ${b}_{H_0,q_0}$, posterior-width ratios ${f}_{H_0,q_0}$ and percentage increase in $H_0$ uncertainty for all combinations of batchsize, learning rate and regularization in the no-selection case, using $[\nt,\nv]=[5000,2000]$.}
\begin{ruledtabular}
\begin{tabular}{@{} cccccccc}
\multicolumn{8}{c}{\textbf{NO SELECTION CASE}} \\ \hline \hline 
 $\nb$ & $\alpha$ & regularizer & $\biasH \;[\hunits]$ & $\biasq$ & $\fsH $ & $\fsq$ & $\incr $ \\
  \hline \hline
  \rowcolor{Gray}
 \multicolumn{8}{c}{\textbf{TRAINING and VALIDATION parameters:} $[\nt,\nv]=[5000,2000]$} \\ \hline \hline
 \multirow{15}{*}{$100$} &  \multirow{5}{*}{$10^{-4}$} & --  & $0.369 \pm 1.752$ & $0.002 \pm 0.098$ & $1.951 \pm 0.124$ & $0.948 \pm 0.023$ & $165.74\%$\\
    & & $\lt = 10^{-4}$  & $-0.002 \pm 0.459$ & $0.008 \pm 0.089$ & $1.055 \pm 0.043$ & $0.95 \pm 0.028$ & $15.59\%$\\
    & & $\lt = 2 \times 10^{-4}$  & $0.054 \pm 0.415$ & $0.008 \pm 0.095$ & $1.056 \pm 0.05$ & $0.95 \pm 0.028$ & $13.95\%$\\
    & & $\lo = 10^{-4}$  & $0.024 \pm 0.35$ & $-0.003 \pm 0.095$ & $1.014 \pm 0.045$ & $0.95 \pm 0.035$ & $7.64\%$\\
    & & $\lo = 2 \times 10^{-4}$  & $0.012 \pm 0.398$ & $0.009 \pm 0.1$ & $1.038 \pm 0.049$ & $0.95 \pm 0.037$ & $11.58\%$\\ \cline{2-8}
    & \multirow{5}{*}{$5 \times 10^{-4}$} & --  & $-0.101 \pm 1.612$ & $-0.003 \pm 0.103$ & $1.855 \pm 0.116$ & $0.948 \pm 0.031$ & $148.89\%$\\
    &  & $\lt = 10^{-4}$  & $0.008 \pm 0.404$ & $0.012 \pm 0.082$ & $1.043 \pm 0.039$ & $0.948 \pm 0.031$ & $12.24\%$\\
    &  & $\lt = 2 \times 10^{-4}$  & $-0.003 \pm 0.423$ & $0.011 \pm 0.084$ & $1.04 \pm 0.042$ & $0.948 \pm 0.029$ & $12.76\%$\\
    &  & $\lo = 10^{-4}$  & $-0.002 \pm 0.365$ & $0.004 \pm 0.098$ & $1.028 \pm 0.042$ & $0.952 \pm 0.032$ & $9.43\%$\\
    &  & $\lo = 2 \times 10^{-4}$  & $-0.006 \pm 0.415$ & $0.011 \pm 0.083$ & $1.027 \pm 0.042$ & $0.947 \pm 0.03$ & $11.27\%$\\ \cline{2-8}
    & \multirow{5}{*}{$10^{-3}$} & --  & $0.053 \pm 0.498$ & $0.005 \pm 0.09$ & $1.058 \pm 0.037$ & $0.954 \pm 0.028$ & $17.55\%$\\
    &  & $\lt = 10^{-4}$  & $-0.014 \pm 0.385$ & $0.009 \pm 0.083$ & $1.04 \pm 0.045$ & $0.952 \pm 0.032$ & $11.32\%$\\
    &  & $\lt = 2 \times 10^{-4}$  & $-0.012 \pm 0.391$ & $0.005 \pm 0.09$ & $1.036 \pm 0.048$ & $0.949 \pm 0.031$ & $11.15\%$\\
    &  & $\lo = 10^{-4}$  & $0.007 \pm 0.352$ & $0.009 \pm 0.09$ & $1.024 \pm 0.048$ & $0.952 \pm 0.038$ & $8.58\%$\\
    &  & $\lo = 2 \times 10^{-4}$  & $-0.026 \pm 0.418$ & $0.011 \pm 0.086$ & $1.022 \pm 0.048$ & $0.948 \pm 0.035$ & $10.94\%$\\ \hline
   \multirow{15}{*}{$500$} & \multirow{5}{*}{$10^{-4}$} & --  & $0.087 \pm 2.066$ & $-0.007 \pm 0.105$ & $2.047 \pm 0.163$ & $0.948 \pm 0.025$ & $195.21\%$\\
    &  & $\lt = 10^{-4}$  & $-0.035 \pm 0.41$ & $0.003 \pm 0.099$ & $1.049 \pm 0.047$ & $0.948 \pm 0.032$ & $13.05\%$\\
    &  & $\lt = 2 \times 10^{-4}$  & $-0.007 \pm 0.375$ & $-0.004 \pm 0.087$ & $1.04 \pm 0.044$ & $0.95 \pm 0.04$ & $10.91\%$\\
    &  & $\lo = 10^{-4}$  & $0.012 \pm 0.358$ & $-0.003 \pm 0.092$ & $1.003 \pm 0.043$ & $0.947 \pm 0.036$ & $6.81\%$\\
    &  & $\lo = 2 \times 10^{-4}$  & $0.01 \pm 0.399$ & $0.002 \pm 0.096$ & $1.041 \pm 0.043$ & $0.948 \pm 0.032$ & $11.94\%$\\ \cline{2-8}
    & \multirow{5}{*}{$5 \times 10^{-4}$} & --  & $-0.195 \pm 1.807$ & $0.003 \pm 0.099$ & $2.068 \pm 0.157$ & $0.948 \pm 0.022$ & $178.17\%$\\
    &  & $\lt = 10^{-4}$  & $0.021 \pm 0.427$ & $0.003 \pm 0.096$ & $1.041 \pm 0.042$ & $0.949 \pm 0.026$ & $13.01\%$\\
    &  & $\lt = 2 \times 10^{-4}$  & $-0.027 \pm 0.388$ & $0.009 \pm 0.099$ & $1.038 \pm 0.053$ & $0.953 \pm 0.035$ & $11.19\%$\\
    &  & $\lo = 10^{-4}$  & $0.026 \pm 0.328$ & $0.001 \pm 0.091$ & $1.018 \pm 0.051$ & $0.948 \pm 0.026$ & $7.3\%$\\
    &  & $\lo = 2 \times 10^{-4}$  & $0.015 \pm 0.361$ & $-0.002 \pm 0.087$ & $1.022 \pm 0.043$ & $0.951 \pm 0.032$ & $8.73\%$\\ \cline{2-8}
    & \multirow{5}{*}{$10^{-3}$} & --  & $0.593 \pm 1.892$ & $0.001 \pm 0.101$ & $2.078 \pm 0.172$ & $0.951 \pm 0.022$ & $184.82\%$\\
    &  & $\lt = 10^{-4}$  & $0.01 \pm 0.413$ & $0.01 \pm 0.088$ & $1.052 \pm 0.044$ & $0.949 \pm 0.034$ & $13.42\%$\\
    &  & $\lt = 2 \times 10^{-4}$  & $0.003 \pm 0.413$ & $0.001 \pm 0.084$ & $1.038 \pm 0.04$ & $0.947 \pm 0.036$ & $12.13\%$\\
    &  & $\lo = 10^{-4}$  & $0.021 \pm 0.322$ & $-0.0 \pm 0.087$ & $1.012 \pm 0.054$ & $0.943 \pm 0.036$ & $6.45\%$\\
    &  & $\lo = 2 \times 10^{-4}$  & $-0.01 \pm 0.362$ & $0.01 \pm 0.101$ & $1.03 \pm 0.053$ & $0.95 \pm 0.041$ & $9.59\%$\\
\end{tabular}
\end{ruledtabular}
\end{table*}
 
\begin{table*}[tp]
\caption{\label{tab:3D_nosel_results_large} Means and standard deviations for the biases ${b}_{H_0,q_0}$, posterior-width ratios ${f}_{H_0,q_0}$ and percentage increase in $H_0$ uncertainty for all combinations of batchsize, learning rate and regularization in the no-selection case, using $[\nt,\nv]=[500000,100000]$.}
\begin{ruledtabular}
\begin{tabular}{@{} cccccccc}
\multicolumn{8}{c}{\textbf{NO SELECTION CASE}} \\ \hline \hline 
 $\nb$ & $\alpha$ & regularizer & $\biasH \;[\hunits]$ & $\biasq$ & $\fsH $ & $\fsq$ & $\incr $ \\
  \hline \hline
 \rowcolor{Gray}
 \multicolumn{8}{c}{\textbf{TRAINING and VALIDATION parameters:} $[\nt,\nv]=[500000,100000]$} \\ \hline \hline 
\multirow{15}{*}{$100$} & \multirow{5}{*}{$10^{-4}$} & --  & $-0.063 \pm 0.253$ & $0.016 \pm 0.065$ & $0.969 \pm 0.042$ & $0.945 \pm 0.038$ & $0.3\%$\\
    &  & $\lt = 10^{-4}$  & $-0.053 \pm 0.193$ & $0.018 \pm 0.065$ & $0.981 \pm 0.047$ & $0.951 \pm 0.042$ & $0.11\%$\\
    &  & $\lt = 2 \times 10^{-4}$  & $-0.073 \pm 0.193$ & $0.015 \pm 0.061$ & $0.979 \pm 0.042$ & $0.945 \pm 0.038$ & $-0.05\%$\\
    &  & $\lo = 10^{-4}$  & $-0.073 \pm 0.243$ & $0.023 \pm 0.062$ & $0.97 \pm 0.044$ & $0.944 \pm 0.038$ & $0.13\%$\\
    &  & $\lo = 2 \times 10^{-4}$  & $-0.075 \pm 0.254$ & $0.006 \pm 0.064$ & $0.975 \pm 0.038$ & $0.943 \pm 0.034$ & $0.93\%$\\ \cline{2-8}
    & \multirow{5}{*}{$5 \times 10^{-4}$} & --  & $-0.061 \pm 0.218$ & $0.014 \pm 0.071$ & $0.978 \pm 0.048$ & $0.948 \pm 0.04$ & $0.35\%$\\
    &  & $\lt = 10^{-4}$  & $-0.058 \pm 0.222$ & $0.015 \pm 0.063$ & $0.972 \pm 0.045$ & $0.945 \pm 0.041$ & $-0.18\%$\\
    &  & $\lt = 2 \times 10^{-4}$  & $-0.044 \pm 0.224$ & $0.016 \pm 0.058$ & $0.981 \pm 0.049$ & $0.945 \pm 0.037$ & $0.79\%$\\
    &  & $\lo = 10^{-4}$  & $-0.032 \pm 0.267$ & $0.009 \pm 0.067$ & $0.968 \pm 0.045$ & $0.94 \pm 0.041$ & $0.57\%$\\
    &  & $\lo = 2 \times 10^{-4}$  & $-0.072 \pm 0.293$ & $0.024 \pm 0.062$ & $0.973 \pm 0.043$ & $0.947 \pm 0.04$ & $1.89\%$\\ \cline{2-8}
    & \multirow{5}{*}{$10^{-3}$} & --  & $-0.058 \pm 0.21$ & $0.02 \pm 0.058$ & $0.973 \pm 0.042$ & $0.945 \pm 0.04$ & $-0.35\%$\\
    &  & $\lt = 10^{-4}$  & $-0.066 \pm 0.224$ & $0.024 \pm 0.063$ & $0.979 \pm 0.044$ & $0.944 \pm 0.035$ & $0.54\%$\\
    &  & $\lt = 2 \times 10^{-4}$  & $-0.039 \pm 0.252$ & $0.022 \pm 0.063$ & $0.979 \pm 0.047$ & $0.946 \pm 0.039$ & $1.23\%$\\
    &  & $\lo = 10^{-4}$  & $-0.074 \pm 0.281$ & $0.021 \pm 0.062$ & $0.98 \pm 0.044$ & $0.947 \pm 0.038$ & $2.14\%$\\
    &  & $\lo = 2 \times 10^{-4}$  & $-0.057 \pm 0.3$ & $0.014 \pm 0.062$ & $0.974 \pm 0.044$ & $0.946 \pm 0.039$ & $2.12\%$\\ \hline
   \multirow{15}{*}{$500$} & \multirow{5}{*}{$10^{-4}$} & --  & $-0.034 \pm 0.264$ & $0.017 \pm 0.065$ & $0.974 \pm 0.044$ & $0.948 \pm 0.042$ & $1.15\%$\\
    &  & $\lt = 10^{-4}$  & $-0.043 \pm 0.193$ & $0.017 \pm 0.066$ & $0.972 \pm 0.041$ & $0.944 \pm 0.039$ & $-0.77\%$\\
    &  & $\lt = 2 \times 10^{-4}$  & $-0.047 \pm 0.196$ & $0.016 \pm 0.059$ & $0.976 \pm 0.044$ & $0.945 \pm 0.034$ & $-0.37\%$\\
    &  & $\lo = 10^{-4}$  & $-0.064 \pm 0.225$ & $0.012 \pm 0.061$ & $0.969 \pm 0.041$ & $0.947 \pm 0.038$ & $-0.41\%$\\
    &  & $\lo = 2 \times 10^{-4}$  & $-0.069 \pm 0.246$ & $0.022 \pm 0.064$ & $0.976 \pm 0.044$ & $0.945 \pm 0.041$ & $0.88\%$\\ \cline{2-8}
    & \multirow{5}{*}{$5 \times 10^{-4}$} & --  & $-0.052 \pm 0.243$ & $0.008 \pm 0.06$ & $0.974 \pm 0.042$ & $0.946 \pm 0.04$ & $0.54\%$\\
    &  & $\lt = 10^{-4}$  & $-0.064 \pm 0.208$ & $0.016 \pm 0.057$ & $0.969 \pm 0.04$ & $0.942 \pm 0.037$ & $-0.79\%$\\
    &  & $\lt = 2 \times 10^{-4}$  & $-0.053 \pm 0.208$ & $0.015 \pm 0.061$ & $0.975 \pm 0.046$ & $0.944 \pm 0.037$ & $-0.15\%$\\
    &  & $\lo = 10^{-4}$  & $-0.056 \pm 0.249$ & $0.022 \pm 0.065$ & $0.967 \pm 0.039$ & $0.944 \pm 0.034$ & $0.0\%$\\
    &  & $\lo = 2 \times 10^{-4}$  & $-0.057 \pm 0.273$ & $0.022 \pm 0.07$ & $0.974 \pm 0.043$ & $0.946 \pm 0.038$ & $1.37\%$\\ \cline{2-8}
    & \multirow{5}{*}{$10^{-3}$} & --  & $-0.056 \pm 0.227$ & $0.019 \pm 0.056$ & $0.975 \pm 0.054$ & $0.944 \pm 0.04$ & $0.3\%$\\
    & & $\lt = 10^{-4}$  & $-0.062 \pm 0.189$ & $0.012 \pm 0.06$ & $0.979 \pm 0.048$ & $0.946 \pm 0.035$ & $-0.22\%$\\
    & & $\lt = 2 \times 10^{-4}$  & $-0.044 \pm 0.229$ & $0.015 \pm 0.066$ & $0.982 \pm 0.045$ & $0.946 \pm 0.04$ & $0.96\%$\\
    & & $\lo = 10^{-4}$  & $-0.076 \pm 0.27$ & $0.014 \pm 0.063$ & $0.97 \pm 0.04$ & $0.946 \pm 0.033$ & $0.88\%$\\
    & & $\lo = 2 \times 10^{-4}$  & $-0.078 \pm 0.285$ & $0.011 \pm 0.064$ & $0.974 \pm 0.041$ & $0.947 \pm 0.037$ & $1.73\%$\\
\end{tabular}
\end{ruledtabular}
\end{table*}

\begin{table*}[htp]
\caption{\label{tab:3D_sel_results} Means and standard deviations for the biases ${b}_{H_0,q_0}$, posterior-width ratios ${f}_{H_0,q_0}$ and percentage increase in $H_0$ uncertainty for all combinations of batchsize, learning rate and regularization in the selection case, using $[\nt,\nv]=[5000,2000]$.}
\begin{ruledtabular}
\begin{tabular}{@{} cccccccc}
\multicolumn{8}{c}{\textbf{SELECTION CASE}} \\ \hline \hline \hspace{0.4cm}
 $\nb$ & $\alpha$ & regularizer & $\biasH \;[\hunits]$ & $\biasq$ & $\fsH $ & $\fsq$ & $\incr $ \\
  \hline \hline
  \rowcolor{Gray}
 \multicolumn{8}{c}{\textbf{TRAINING and VALIDATION parameters:} $[\nt,\nv]=[5000,2000]$} \\ \hline \hline
    \multirow{15}{*}{$100$} & \multirow{5}{*}{$10^{-4}$} & --  & $-0.153 \pm 1.714$ & $0.014 \pm 0.136$ & $1.77 \pm 0.139$ & $1.019 \pm 0.076$ & $144.02\%$\\
    &  & $\lt = 10^{-4}$  & $0.043 \pm 0.403$ & $0.02 \pm 0.123$ & $1.059 \pm 0.044$ & $1.012 \pm 0.04$ & $13.05\%$\\
    &  & $\lt = 2 \times 10^{-4}$  & $0.021 \pm 0.426$ & $0.016 \pm 0.114$ & $1.047 \pm 0.037$ & $1.013 \pm 0.044$ & $12.74\%$\\
    &  & $\lo = 10^{-4}$  & $-0.015 \pm 0.338$ & $0.014 \pm 0.115$ & $1.013 \pm 0.039$ & $1.005 \pm 0.044$ & $6.53\%$\\
    &  & $\lo = 2 \times 10^{-4}$  & $-0.029 \pm 0.365$ & $0.008 \pm 0.119$ & $1.029 \pm 0.038$ & $1.008 \pm 0.044$ & $8.97\%$\\ \cline{2-8}
    & \multirow{5}{*}{$5 \times 10^{-4}$} & --  & $-0.309 \pm 1.657$ & $0.007 \pm 0.126$ & $1.838 \pm 0.153$ & $1.013 \pm 0.032$ & $145.28\%$\\
    &  & $\lt = 10^{-4}$  & $0.059 \pm 0.423$ & $0.009 \pm 0.116$ & $1.018 \pm 0.043$ & $1.007 \pm 0.057$ & $9.95\%$\\
    &  & $\lt = 2 \times 10^{-4}$  & $-0.015 \pm 0.4$ & $0.022 \pm 0.118$ & $1.033 \pm 0.041$ & $1.009 \pm 0.044$ & $10.47\%$\\
    &  & $\lo = 10^{-4}$  & $-0.02 \pm 0.313$ & $-0.001 \pm 0.122$ & $1.014 \pm 0.041$ & $1.005 \pm 0.058$ & $5.9\%$\\
    &  & $\lo = 2 \times 10^{-4}$  & $0.0 \pm 0.393$ & $0.003 \pm 0.115$ & $1.036 \pm 0.038$ & $1.009 \pm 0.042$ & $10.5\%$\\ \cline{2-8}
    & \multirow{5}{*}{$10^{-3}$} & --  & $-0.006 \pm 0.526$ & $0.011 \pm 0.132$ & $1.055 \pm 0.055$ & $1.015 \pm 0.089$ & $17.38\%$\\
    &  & $\lt = 10^{-4}$  & $-0.03 \pm 0.408$ & $0.011 \pm 0.114$ & $1.013 \pm 0.044$ & $1.005 \pm 0.056$ & $8.93\%$\\
    &  & $\lt = 2 \times 10^{-4}$  & $0.017 \pm 0.387$ & $0.02 \pm 0.121$ & $1.023 \pm 0.046$ & $1.009 \pm 0.063$ & $9.09\%$\\
    &  & $\lo = 10^{-4}$  & $0.014 \pm 0.357$ & $0.008 \pm 0.119$ & $1.018 \pm 0.042$ & $1.006 \pm 0.051$ & $7.67\%$\\
    &  & $\lo = 2 \times 10^{-4}$  & $-0.003 \pm 0.476$ & $0.0 \pm 0.12$ & $1.055 \pm 0.048$ & $1.011 \pm 0.056$ & $15.32\%$\\ \cline{2-8}
   \multirow{15}{*}{$500$} & \multirow{5}{*}{$10^{-4}$} & --  & $0.022 \pm 1.961$ & $0.005 \pm 0.136$ & $2.028 \pm 0.176$ & $1.005 \pm 0.028$ & $179.37\%$\\
    &  & $\lt = 10^{-4}$  & $0.011 \pm 0.455$ & $0.016 \pm 0.119$ & $1.044 \pm 0.035$ & $1.009 \pm 0.051$ & $13.52\%$\\
    &  & $\lt = 2 \times 10^{-4}$  & $-0.009 \pm 0.417$ & $0.011 \pm 0.114$ & $1.056 \pm 0.04$ & $1.008 \pm 0.048$ & $13.25\%$\\
    &  & $\lo = 10^{-4}$  & $-0.002 \pm 0.334$ & $0.019 \pm 0.116$ & $1.025 \pm 0.04$ & $1.01 \pm 0.049$ & $7.63\%$\\
    &  & $\lo = 2 \times 10^{-4}$  & $-0.018 \pm 0.377$ & $0.013 \pm 0.125$ & $1.033 \pm 0.037$ & $1.016 \pm 0.03$ & $9.66\%$\\ \cline{2-8}
    & \multirow{5}{*}{$5 \times 10^{-4}$} & --  & $0.189 \pm 2.115$ & $0.027 \pm 0.127$ & $1.97 \pm 0.134$ & $1.017 \pm 0.042$ & $185.92\%$\\
    &  & $\lt = 10^{-4}$  & $-0.039 \pm 0.453$ & $0.02 \pm 0.12$ & $1.048 \pm 0.039$ & $1.012 \pm 0.056$ & $13.75\%$\\
    &  & $\lt = 2 \times 10^{-4}$  & $0.011 \pm 0.408$ & $0.015 \pm 0.119$ & $1.039 \pm 0.035$ & $1.01 \pm 0.046$ & $11.34\%$\\
    &  & $\lo = 10^{-4}$  & $0.001 \pm 0.313$ & $0.012 \pm 0.128$ & $1.025 \pm 0.038$ & $1.013 \pm 0.036$ & $6.99\%$\\
    &  & $\lo = 2 \times 10^{-4}$  & $-0.037 \pm 0.359$ & $0.014 \pm 0.12$ & $1.038 \pm 0.044$ & $1.01 \pm 0.058$ & $9.62\%$\\ \cline{2-8}
    & \multirow{5}{*}{$10^{-3}$} & --  & $-0.082 \pm 1.972$ & $0.016 \pm 0.117$ & $2.027 \pm 0.173$ & $1.012 \pm 0.031$ & $180.05\%$\\
    &  & $\lt = 10^{-4}$  & $-0.006 \pm 0.451$ & $0.011 \pm 0.119$ & $1.046 \pm 0.036$ & $1.005 \pm 0.062$ & $13.57\%$\\
    &  & $\lt = 2 \times 10^{-4}$  & $-0.027 \pm 0.371$ & $0.017 \pm 0.124$ & $1.046 \pm 0.042$ & $1.005 \pm 0.058$ & $10.71\%$\\
    &  & $\lo = 10^{-4}$  & $0.051 \pm 0.329$ & $0.011 \pm 0.137$ & $1.019 \pm 0.053$ & $1.011 \pm 0.058$ & $6.91\%$\\
    &  & $\lo = 2 \times 10^{-4}$  & $0.004 \pm 0.348$ & $0.011 \pm 0.113$ & $1.027 \pm 0.038$ & $1.003 \pm 0.05$ & $8.2\%$\\
\end{tabular}
\end{ruledtabular}
\end{table*}
 
\begin{table*}[tp]
\caption{\label{tab:3D_sel_results_large} Means and standard deviations for the biases ${b}_{H_0,q_0}$, posterior-width ratios ${f}_{H_0,q_0}$ and percentage increase in $H_0$ uncertainty for all combinations of batchsize, learning rate and regularization in the selection case, using $[\nt,\nv]=[500000,100000]$.}
\begin{ruledtabular}
\begin{tabular}{@{} cccccccc}
\multicolumn{8}{c}{\textbf{SELECTION CASE}} \\ \hline \hline 
 $\nb$ & $\alpha$ & regularizer & $\biasH \;[\hunits]$ & $\biasq$ & $\fsH $ & $\fsq$ & $\incr $ \\
  \hline \hline
 \rowcolor{Gray}
 \multicolumn{8}{c}{\textbf{TRAINING and VALIDATION parameters:} $[\nt,\nv]=[500000,100000]$} \\ \hline \hline 
 \multirow{15}{*}{$100$} & \multirow{5}{*}{$10^{-4}$} & --  & $-0.033 \pm 0.278$ & $0.023 \pm 0.082$ & $0.97 \pm 0.037$ & $0.999 \pm 0.045$ & $0.78\%$\\
    &  & $\lt = 10^{-4}$  & $-0.032 \pm 0.184$ & $0.022 \pm 0.092$ & $0.976 \pm 0.031$ & $1.006 \pm 0.036$ & $-0.73\%$\\
    &  & $\lt = 2 \times 10^{-4}$  & $-0.019 \pm 0.195$ & $0.017 \pm 0.085$ & $0.981 \pm 0.036$ & $1.002 \pm 0.037$ & $-0.04\%$\\
    &  & $\lo = 10^{-4}$  & $-0.025 \pm 0.186$ & $0.021 \pm 0.085$ & $0.978 \pm 0.033$ & $1.003 \pm 0.039$ & $-0.55\%$\\
    &  & $\lo = 2 \times 10^{-4}$  & $-0.044 \pm 0.214$ & $0.018 \pm 0.087$ & $0.984 \pm 0.036$ & $1.005 \pm 0.042$ & $0.57\%$\\ \cline{2-8}
    & \multirow{5}{*}{$5 \times 10^{-4}$} & --  & $0.01 \pm 0.207$ & $0.02 \pm 0.087$ & $0.968 \pm 0.038$ & $1.001 \pm 0.04$ & $-1.12\%$\\
    &  & $\lt = 10^{-4}$  & $-0.033 \pm 0.177$ & $0.02 \pm 0.092$ & $0.979 \pm 0.039$ & $1.003 \pm 0.043$ & $-0.56\%$\\
    &  & $\lt = 2 \times 10^{-4}$  & $-0.026 \pm 0.199$ & $0.015 \pm 0.088$ & $0.979 \pm 0.037$ & $1.002 \pm 0.04$ & $-0.21\%$\\
    &  & $\lo = 10^{-4}$  & $-0.028 \pm 0.198$ & $0.019 \pm 0.081$ & $0.988 \pm 0.037$ & $1.001 \pm 0.042$ & $0.64\%$\\
    &  & $\lo = 2 \times 10^{-4}$  & $-0.047 \pm 0.269$ & $0.018 \pm 0.084$ & $0.989 \pm 0.034$ & $1.0 \pm 0.045$ & $2.37\%$\\ \cline{2-8}
    & \multirow{5}{*}{$10^{-3}$} & --  & $0.0 \pm 0.183$ & $0.026 \pm 0.091$ & $0.965 \pm 0.03$ & $1.004 \pm 0.038$ & $-1.88\%$\\
    & & $\lt = 10^{-4}$  & $-0.007 \pm 0.184$ & $0.014 \pm 0.088$ & $0.98 \pm 0.034$ & $1.005 \pm 0.044$ & $-0.35\%$\\
    & & $\lt = 2 \times 10^{-4}$  & $-0.015 \pm 0.193$ & $0.015 \pm 0.093$ & $0.982 \pm 0.037$ & $1.004 \pm 0.044$ & $-0.0\%$\\
    & & $\lo = 10^{-4}$  & $-0.053 \pm 0.242$ & $0.015 \pm 0.087$ & $0.996 \pm 0.033$ & $1.001 \pm 0.045$ & $2.39\%$\\
    & & $\lo = 2 \times 10^{-4}$  & $-0.031 \pm 0.263$ & $0.015 \pm 0.095$ & $0.991 \pm 0.036$ & $1.001 \pm 0.048$ & $2.44\%$\\ \cline{2-8}
   \multirow{15}{*}{$500$} & \multirow{5}{*}{$10^{-4}$} & --  & $-0.037 \pm 0.267$ & $0.022 \pm 0.084$ & $0.98 \pm 0.041$ & $0.998 \pm 0.036$ & $1.42\%$\\
    &  & $\lt = 10^{-4}$  & $-0.038 \pm 0.199$ & $0.028 \pm 0.109$ & $0.976 \pm 0.034$ & $1.01 \pm 0.035$ & $-0.51\%$\\
    &  & $\lt = 2 \times 10^{-4}$  & $-0.02 \pm 0.194$ & $0.021 \pm 0.095$ & $0.971 \pm 0.034$ & $1.005 \pm 0.036$ & $-1.05\%$\\
    &  & $\lo = 10^{-4}$  & $-0.022 \pm 0.178$ & $0.015 \pm 0.093$ & $0.978 \pm 0.036$ & $1.003 \pm 0.039$ & $-0.7\%$\\
    &  & $\lo = 2 \times 10^{-4}$  & $-0.025 \pm 0.186$ & $0.012 \pm 0.089$ & $0.979 \pm 0.036$ & $1.002 \pm 0.035$ & $-0.47\%$\\ \cline{2-8}
    & \multirow{5}{*}{$5 \times 10^{-4}$} & --  & $-0.045 \pm 0.277$ & $0.019 \pm 0.09$ & $0.982 \pm 0.039$ & $1.002 \pm 0.037$ & $1.83\%$\\
    &  & $\lt = 10^{-4}$  & $-0.013 \pm 0.18$ & $0.019 \pm 0.086$ & $0.977 \pm 0.035$ & $1.006 \pm 0.038$ & $-0.68\%$\\
    &  & $\lt = 2 \times 10^{-4}$  & $-0.02 \pm 0.182$ & $0.014 \pm 0.087$ & $0.981 \pm 0.032$ & $1.005 \pm 0.039$ & $-0.32\%$\\
    &  & $\lo = 10^{-4}$  & $-0.03 \pm 0.196$ & $0.018 \pm 0.079$ & $0.982 \pm 0.032$ & $1.001 \pm 0.036$ & $0.04\%$\\
    &  & $\lo = 2 \times 10^{-4}$  & $-0.047 \pm 0.233$ & $0.012 \pm 0.089$ & $0.987 \pm 0.038$ & $1.004 \pm 0.036$ & $1.33\%$\\ \cline{2-8}
    & \multirow{5}{*}{$10^{-3}$} & --  & $-0.013 \pm 0.22$ & $0.021 \pm 0.089$ & $0.962 \pm 0.042$ & $1.003 \pm 0.044$ & $-1.43\%$\\
    &  & $\lt = 10^{-4}$  & $-0.006 \pm 0.201$ & $0.015 \pm 0.086$ & $0.98 \pm 0.033$ & $1.004 \pm 0.041$ & $-0.07\%$\\
    &  & $\lt = 2 \times 10^{-4}$  & $-0.01 \pm 0.199$ & $0.021 \pm 0.083$ & $0.979 \pm 0.043$ & $1.003 \pm 0.042$ & $-0.18\%$\\
    &  & $\lo = 10^{-4}$  & $-0.026 \pm 0.212$ & $0.015 \pm 0.084$ & $0.986 \pm 0.037$ & $1.0 \pm 0.044$ & $0.77\%$\\
    &  & $\lo = 2 \times 10^{-4}$  & $-0.036 \pm 0.25$ & $0.012 \pm 0.09$ & $0.993 \pm 0.037$ & $1.005 \pm 0.043$ & $2.28\%$\\
\end{tabular}
\end{ruledtabular}
\end{table*}

\begin{figure*}
    \centering
    \includegraphics[width=\textwidth]{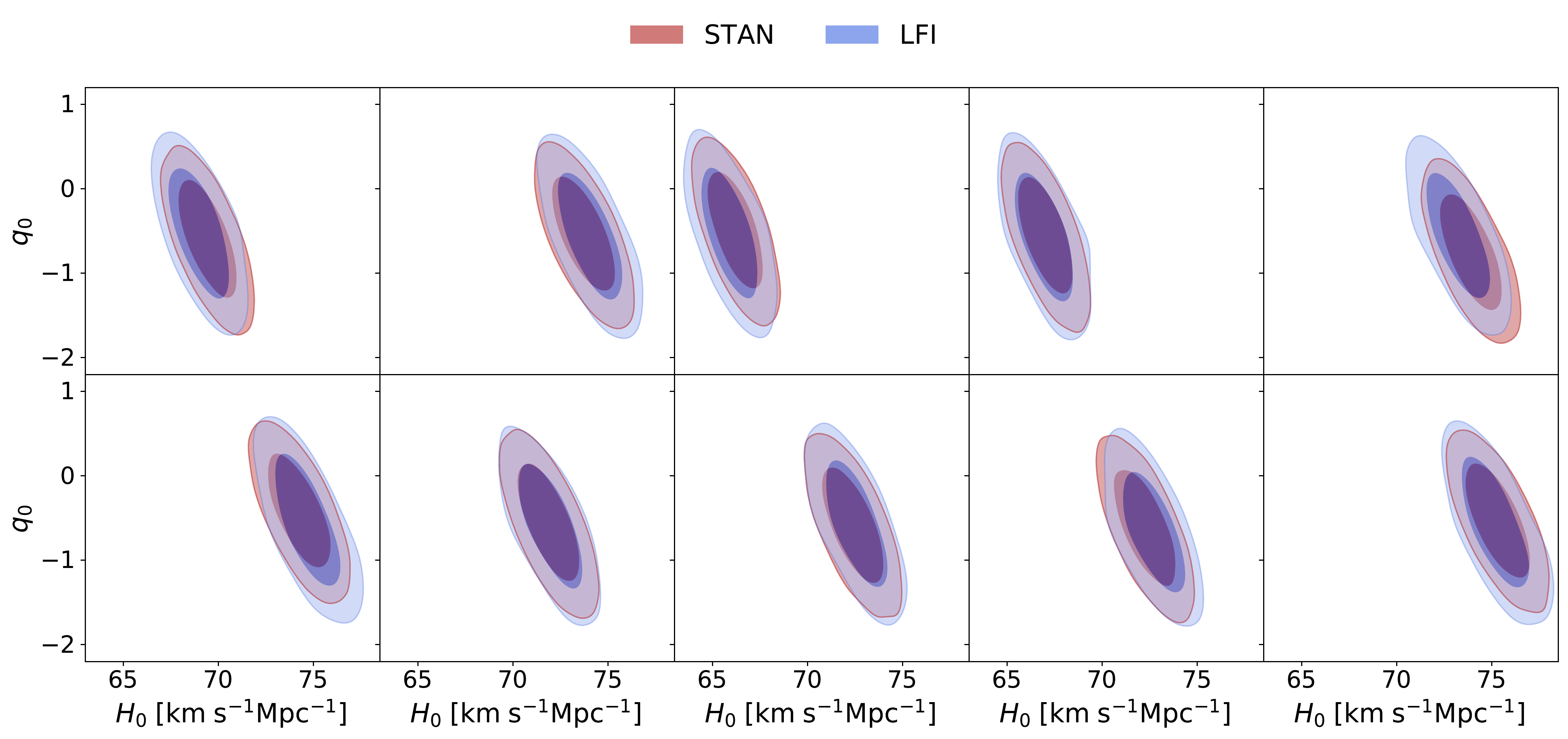}
    \caption{Example posterior contour plots produced by LFI (blue) and traditional Bayesian sampling (red) for test datasets with GW selection.} 
    \label{fig:posteriorplots}
\end{figure*}

\end{document}